\documentclass[12pt]{article}
\pdfoutput=1

\usepackage{pgfplots}
\usetikzlibrary{decorations.pathmorphing}

\tikzset{snake it/.style={decorate, decoration=snake}}
\pgfplotsset{compat=1.10}
\usepgfplotslibrary{fillbetween}
\usetikzlibrary{patterns}

\usepackage{draft} 
\usepackage{hyperref}
\usepackage{graphicx,color,subfig}
\usepackage{cite}
\usepackage{mciteplus}
\usepackage{skak}
\usepackage{empheq}
\usepackage{tikz}
\DeclareFontFamily{OT1}{pzc}{}
\DeclareFontShape{OT1}{pzc}{m}{it}{<-> s * [1.10] pzcmi7t}{}
\DeclareMathAlphabet{\mathpzc}{OT1}{pzc}{m}{it}

\def\be#1\ee{\begin{align}#1\end{align}}

\makeatletter
\newcommand{\bdryno}{\mathpalette\bdry@no\relax}
\newcommand{\bdry@no}[2]{%
  \mspace{1mu}%
  \vbox{%
    \hbox{$\m@th#1\scriptstyle{\ast}$}
    \nointerlineskip
    \kern.25ex
    \hbox{$\m@th#1\scriptstyle{\ast}$}
    \kern-.06ex
  }%
  \mspace{1mu}%
}
\makeatother

\newcommand{\bno}{\,\bdryno\,}

\usetikzlibrary{shapes.misc}
\tikzset{cross/.style={cross out, draw=black, minimum size=2*(#1-\pgflinewidth), inner sep=0pt, outer sep=0pt},
cross/.default={1pt}}

\begin{document}

\unitlength = .8mm

\begin{titlepage}

\begin{center}

\hfill \\
\hfill \\
\vskip 1cm

\title{Long String Scattering in $c=1$ String Theory}

\author{Bruno Balthazar$^\spadesuit$, Victor A. Rodriguez$^\spadesuit$, Xi Yin$^\spadesuit{}^\diamondsuit$}

\address{
$^\spadesuit$Jefferson Physical Laboratory, Harvard University, \\
Cambridge, MA 02138 USA
\\
$^\diamondsuit$Center for Theoretical Physics, Massachusetts Institute of Technology, \\
Cambridge, MA 02139 USA
}

\email{bbalthazar@g.harvard.edu, victorrodriguez@g.harvard.edu, xiyin@g.harvard.edu}

\end{center}

\abstract{We study the scattering of long strings in $c=1$ string theory, both in the worldsheet description and in the non-singlet sector of the dual matrix quantum mechanics. From the worldsheet perspective, the scattering amplitudes of long strings are obtained from a decoupling limit of open strings amplitudes on FZZT branes, which we compute by integrating Virasoro conformal blocks along with structure constants of boundary Liouville theory. In particular, we study the tree level amplitudes of (1) a long string decaying by emitting a closed string, and (2) the scattering of a pair of long strings. We show that they are indeed well defined as limits of open string amplitudes, and that our results are in striking numerical agreement with computations in the adjoint and bi-adjoint sectors of the dual matrix model (based on proposals of Maldacena and solutions due to Fidkowski), thereby providing strong evidence of the duality.}

\vfill

\end{titlepage}

\eject

\begingroup
\hypersetup{linkcolor=black}
\tableofcontents
\endgroup

\section{Introduction} 

The $c=1$ string theory is defined perturbatively through the worldsheet CFT of time-like free boson $X^0$, $c=25$ Liouville theory, together with the $b, c$ conformal ghosts. The physical closed string degrees of freedom in the target spacetime are those of a single scalar field in 1+1 dimensions, which is conjectured to be dual to the collective excitations of the fermi surface in the system of a large number of free fermions governed by a certain non-relativistic Hamiltonian \cite{Klebanov:1991qa, Ginsparg:1993is, Jevicki:1993qn, Polchinski:1994mb, Martinec:2004td}. The fermion system is equivalent to a suitable large $N$ scaling limit of a $U(N)$ gauged Hermitian matrix model, which we refer to as the $c=1$ matrix model. Early investigations of $c=1$ string theory/matrix model duality preceded the exact solution of Liouville CFT \cite{Dorn:1994xn, Zamolodchikov:1995aa, Teschner:1995yf, Teschner:2001rv}. The S-matrix of closed strings in $c=1$ string theory has been re-analyzed recently in \cite{Balthazar:2017mxh}, using Liouville correlation functions evaluated by integrating Virasoro conformal blocks, and was found to agree with that of the matrix model for genus zero 4-point amplitudes and genus one 2-point amplitudes, giving highly nontrivial support to the conjectured duality.

The $c=1$ string theory further admits two types of D-branes, namely the ZZ-brane \cite{Zamolodchikov:2001ah} and the FZZT-branes \cite{Fateev:2000ik, Teschner:2000md}. Let $\phi$ be the target space coordinate parameterizing the Liouville direction. The ZZ-brane can be thought of as a 0-brane that is localized in the strong coupling region (positive $\phi$), and is unstable due to an open string tachyon mode. The FZZT-branes, on the other hand, are a one-parameter family of 1-branes that extend from the weak coupling region $\phi\to -\infty$ to a finite value of $\phi$. An FZZT brane supports open string states that behave in the weak coupling region as modes of a single massless scalar field.

The ZZ-brane was proposed in \cite{McGreevy:2003kb} to be dual to a single fermion, or eigenvalue, of the dual Hermitian matrix model. The role of the FZZT brane in the matrix model was suggested in \cite{Gaiotto:2005gd}, but its precise matrix model description remains to be understood. An interesting limit was considered in \cite{Maldacena:2005hi}, in which the FZZT brane recedes to the weak coupling region $\phi\to -\infty$, while a high energy open string mode on the FZZT brane moves toward the strong coupling region and stretches to some finite value of $\phi$, before retracting and moving back to $\phi\to -\infty$. Such an open string state, known as the {\it long string}, was conjectured to be dual to a state in the adjoint sector of the Hermitian matrix model \cite{Maldacena:2005hi}.

The aim of this paper is to explore the conjecture of \cite{Maldacena:2005hi} by a detailed comparison of tree level scattering amplitudes involving the long strings and closed strings in $c=1$ string theory with those of the matrix model, focusing on two examples: (1) a long string decaying into a long and a closed string, and (2) the scattering of a pair of long strings. From the worldsheet CFT, the relevant amplitude can be evaluated by integrating suitable correlators of Liouville theory on the disc, subject to FZZT boundary conditions, in the conformally invariant cross ratio. The disc correlators in question are computed by integrating boundary/bulk structure functions of Liouville theory against the appropriate Virasoro conformal blocks, and then taking the long string limit. As a preliminary but nontrivial check, we will see that the disc amplitudes of the long strings are indeed finite and well defined.

\begin{figure}[h!]
\centering
\subfloat[]{~~~~~
\begin{tikzpicture}
\filldraw[color=black, fill=black!10, thick] (0,0) circle (2);
\draw (0,0) node[cross=4pt, very thick] {};
\draw (2,0) node[cross=4pt, very thick] {};
\draw (-2,0) node[cross=4pt, very thick] {};
\draw (0,0) node[above] {$C$, out};
\draw (2,0) node[right] {$L$, out};
\draw (-2,0) node[left] {$L$, in};
\end{tikzpicture}~~~~~
}~~
\subfloat[]{~~~~~
\begin{tikzpicture}
\filldraw[color=black, fill=black!10, thick] (0,0) circle (2);
\draw (2,0) node[cross=4pt, very thick] {};
\draw (-2,0) node[cross=4pt, very thick] {};
\draw (0,-2) node[cross=4pt, very thick] {};
\draw (0,2) node[cross=4pt, very thick] {};
\draw (2,0) node[right] {$L$, out};
\draw (-2,0) node[left] {$L$, out};
\draw (0,-2) node[below] {$L$, in};
\draw (0,2) node[above] {$L$, in};
\end{tikzpicture}~~~~~
}
\caption{(a) Worldsheet diagram of closed string emission by a long string at tree level. (b) Worldsheet diagram for the scattering of a pair of long strings at tree level.}
\end{figure}
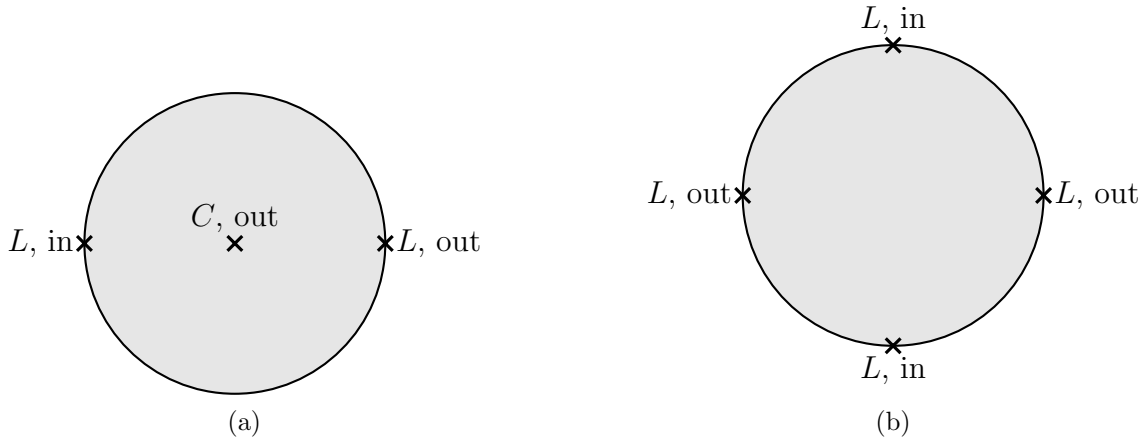

We then consider the scattering amplitudes in the adjoint and bi-adjoint sectors of the matrix model, corresponding to one and two long strings respectively. The long string asymptotic states are constructed following the prescription of\cite{Maldacena:2005hi} and making use of the exact solution of \cite{Fidkowski:2005ck} 
The closed strings are treated as collective excitations of the density of matrix eigenvalues on top of the long string state. After a careful rewriting of the Hamiltonian in terms of integrals over the eigenvalue distribution, the tree level amplitudes can be computed from the Born approximation and evaluated numerically as a function of the closed string energy and the renormalized long string energies. We find striking agreement with the worldsheet results, for both the long $\to$ long $+$ closed string amplitude and the long $+$ long $\to$ long $+$ long string amplitude, to good numerical precision, thereby providing a strong check of the proposed duality.


The paper is organized as follows. In section \ref{sec:LiouvCFT} we will review Liouville CFT with and without boundaries, and define the long string limit. In section 3 the tree-level worldsheet scattering amplitudes for long $\to$ long $+$ closed and long $+$ long $\to$ long $+$ long are computed. In section \ref{sec:MM} we will describe the non-singlet sector of the $c=1$ matrix model. We will review the fermi sea description of the matrix model eigenvalues and the long string wavefunctions. Finally, we will numerically compute scattering amplitudes in the adjoint and bi-adjoint sector and obtain exact numerical agreement with the worldsheet scattering amplitudes described above. 

We will conclude and discuss future directions in section \ref{sec:conc}. Details of the crossing conditions in boundary Liouville theory, further numerical details of the computations of scattering amplitudes, and other minor results are presented in the appendices.

\section{FZZT branes and long strings in $c=1$ string theory}
\label{sec:LiouvCFT}

\subsection{The closed string sector of $c=1$ string theory}
\label{closedreview}

We begin by reviewing the closed string sector of $c=1$ string theory, in the absence of branes. The worldsheet CFT consists of a timelike free boson $X^0$, a $c=25$ Liouville theory, and the $b,c$ conformal ghosts. The Liouville theory can be described by the action
\ie
S_L[\phi] = {1\over 4\pi} \int d^2\sigma \sqrt{g} \left( g^{mn} \partial_m \phi \partial_n \phi + Q R \phi + 4\pi \mu e^{2b\phi} \right)
\fe 
in the semi-classical regime, where the background charge $Q=b+b^{-1}$ is related to the central charge $c$ via $c=1+6Q^2$. The case of interest, namely $c=25$, corresponds to $b=1$.

The complete set of Virasoro primaries in the $c=25$ Liouville CFT are given by scalar operators $V_P$, of conformal weight $h=\tilde h = 1+P^2$, where $P\in\mathbb{R}_{\geq 0}$ is the ``Liouville momentum". Our convention is such that $V_P$ are delta-function normalized, namely their two-point functions take the form
\ie
\left\langle V_{P}(z,\bar z) V_{P'}(0) \right\rangle = \pi{\delta(P-P')\over |z|^{4h}}.
\fe
In the ``weak coupling" regime $\phi\to-\infty$, $V_P$ admits a free field representation, in the form of a reflection wave
\ie
V_P\sim S(P)^{-\frac{1}{2}}e^{(2+2i P)\phi}+S(P)^{\frac{1}{2}}e^{(2-2i P)\phi}, 
\fe
where the reflection phase $S(P)$ is given by
\ie
S(P)=-\left(\frac{\Gamma(2 i P)}{\Gamma(-2 i P)}\right)^2.
\fe
The 3-point function coefficients of Liouville theory, known as DOZZ structure constants \cite{Dorn:1994xn, Zamolodchikov:1995aa}, are given in the $c=25$ case by 
\ie\label{cppp}
{\cal C}(P_1, P_2, P_3) = {1\over \Upsilon_1(1+i(P_1+P_2+P_3)) } \left[ { 2P_1 \Upsilon_1(1+2iP_1)\over  \Upsilon_1(1 + i(P_2+P_3-P_1)) } \times (2~{\rm permutations}) \right].
\fe
Here $\Upsilon_1(x)$ is a special case of Barnes double Gamma function, defined as 
\ie
\Upsilon_1(x)={1\over {\Gamma_1(x)\Gamma_1(2-x)}},
\fe
where the function $\Gamma_1(x)$ is related to the Barnes G-function $G(x)$ by $\Gamma_1(x)={(2\pi)^{(x-1)/2}}(G(x))^{-1}$. Importantly, $\Gamma_1(x)$ is a meromorphic function with (not necessarily simple) poles at $x\in\mathbb{Z}_{\leq0}$, and obeys the recursion relation
\ie
\Gamma_1(x+1)=\frac{\sqrt{2\pi}}{\Gamma(x)}\Gamma_1(x).
\fe
Consequently, $\Upsilon_1(x)$  is an entire analytic function with zeroes at $x\in\mathbb{Z}\setminus \{1\}$, and obeys
\ie
\Upsilon_1(2-x) = \Upsilon_1(x),
~~~~ \Upsilon_1(x+1) = \frac{\Gamma(x)}{\Gamma(1-x)} \Upsilon_1(x).
\fe

The closed string asymptotic states are given by BRST cohomology classes represented by vertex operators of the form
\ie\label{ima}
{\cal V}_\omega^\pm = g_s :\! e^{\pm i\omega X^0}\!\!: V_{P={\omega\over 2}}.
\fe
Here ${\cal V}_\omega^+$ represents an in-state and ${\cal V}_\omega^-$ an out-state, of energy $\omega\geq 0$, normalized according to 
\ie\label{closednormal}
\langle\omega|\omega'\rangle = \omega \delta(\omega-\omega').
\fe 
The perturbative amplitudes of the closed strings are studied in \cite{Balthazar:2017mxh} explicitly at genus zero and one, and were found to be in agreement with computations in the dual matrix quantum mechanics.

\subsection{FZZT branes, open strings, and the long string limit}
\label{fzztbraneintro}

The Liouville CFT admits two types of unitary conformal boundary conditions: the ZZ boundary condition \cite{Zamolodchikov:2001ah} corresponding to a boundary state $|{\rm ZZ}\rangle$, and the FZZT boundary condition \cite{Fateev:2000ik} described by a 1-parameter family of boundary states $|{\rm FZZT}(s)\rangle$. In $c=1$ string theory, a corresponding brane is defined on the worldsheet through a conformal boundary condition that is Neumann in $X^0$ and of ZZ or FZZT type in the Liouville CFT.

The ZZ boundary state takes the form
\ie
|{\rm ZZ}\rangle=\int_0^\infty \frac{dP}{\pi}\, \Psi^{\rm ZZ}(P) |V_P\rangle\rangle,
\fe
where $|V_P\rangle\rangle$ is the Ishibashi state constructed from the primary $V_P$, and $\Psi^{\rm ZZ}(P)$ is given by\footnote{In the language of \cite{Zamolodchikov:2001ah}, $|{\rm ZZ}\rangle$ is the ZZ boundary state of type $(1,1)$, which is the only admissible unitary boundary condition in Liouville CFT with $c>1$.} 
\ie\label{eq:ZZ1pt}
\Psi^{\rm ZZ}(P) = 2^{5\over 4}\sqrt{\pi} \sinh(2\pi P).
\fe
The ZZ boundary condition admits only one boundary Virasoro primary, namely the identity operator. 
Indeed, the cylinder partition function of Liouville CFT subject to ZZ boundary condition on both boundaries is given by
\ie\label{cardyzz}
\int_{0}^{\infty}\frac{dP}{\pi} \left(\Psi^{\rm ZZ}(P)\right)^2\chi_{1+P^2}(\tau) = \widehat\chi_0(-1/\tau),
\fe
where $\chi_h(\tau)$ is the $c=25$ Virasoro character of a primary of weight $h$, and $\widehat\chi_0(\tau)$ is the degenerate vacuum character.
This gives rise to a tachyonic open string mode that renders the ZZ-brane unstable. The role of the ZZ-brane in $c=1$ string theory was pointed out in \cite{McGreevy:2003kb}, but is not of concern in this paper.

The FZZT boundary state takes the form
\ie
|{\rm FZZT}(s)\rangle=\int_0^\infty \frac{dP}{\pi}\, \Psi^{\rm FZZT}_s(P) |V_P\rangle\rangle,
\fe
where
\ie\label{eq:FZZ1pt}
\Psi^{\rm FZZT}_s(P) = 2^{1\over 4} \sqrt{\pi}{\cos(4\pi s P) \over \sinh(2\pi P)}.
\fe
It is such that the Liouville CFT on the strip with ZZ boundary condition on one side and FZZT boundary condition on the other admits a unique Virasoro primary, namely ${\cal H}_{{\rm ZZ, FZZT}(s)}$ is spanned by the Virasoro descendants of a single boundary primary, of conformal weight $1+s^2$. This can be seen from the cylinder partition function
\ie\label{cardyzzfzzt}
\int_{0}^{\infty}\frac{dP}{\pi}\, \Psi^{\rm ZZ}(P)\Psi^{\rm FZZT}_s(P)\chi_{1+P^2}(\tau)= \chi_{1+s^2}(-1/\tau).
\fe
The FZZT boundary condition is unitary provided that $s$ is either a non-negative real number or purely imaginary with $0\leq {\rm Im} s<1$. 

The open string spectrum on the FZZT brane is such that the FZZT brane is stable for either non-negative real $s$ or for $s$ purely imaginary subject to $0\leq {\rm Im} s<{1\over 2}$. For $s_1$ and $s_2$ in this regime, the Hilbert space on the strip with FZZT boundary conditions specified by $s_1$ on one side and $s_2$ on the other,  ${\cal H}_{{\rm FZZT}(s_1), {\rm FZZT}(s_2)}$, consists of a continuous family of boundary Virasoro primaries $\psi_P^{s_1, s_2}$ labeled by ``Liouville momentum" $P\in \mathbb{R}_{\geq 0}$, of conformal weight $h_P = 1+P^2$. 

In this paper, we are interested only in the stable FZZT branes, particularly in the large $s$ limit. Heuristically, one may think of the ZZ brane as a 0-brane localized in the strong coupling regime ($\phi \sim 0$), whereas the FZZT brane is a sort of 1-brane that extends from the weak coupling limit $\phi\to -\infty$ to $\phi\sim -s$, so that the open string stretched between a ZZ and a FZZT$(s)$ brane has mass $\sim s$.

%

Let us comment that the FZZT boundary condition admits a Lagrangian description in the semi-classical regime, as a Neumann boundary condition on the Liouville field $\phi$, with bulk-boundary action
\ie
S_L[\phi] = {1\over 4\pi} \int_\Sigma d^2\sigma \sqrt{g} \left( g^{mn} \partial_m \phi \partial_n \phi + Q R \phi + 4\pi \mu e^{2b\phi} \right)+\int_{\partial\Sigma}d\xi g^{\frac{1}{4}}\left(\frac{Qk}{2\pi}\phi+\mu_B e^{b\phi}\right),
\fe 
where $d\xi g^{1\over 4}$ is the boundary line element and $k$ is the extrinsic curvature of the boundary. The parameter $\mu_B$, known as the boundary cosmological constant, is related to the parameter $s$ via
\ie
\cosh(2\pi s)=\frac{\mu_B}{\sqrt{\mu}}\sqrt{\sin(\pi b^2)}.
\fe
Note that this relation is singular in the $b\to 1$ limit, where the semi-classical description of the FZZT boundary condition breaks down. As this is the case of interest in $c=1$ string theory, we will exclusively work with $s$, rather than $\mu_B$, as the true parameter of FZZT branes.



We will normalize the boundary primaries according to the disc 2-point function
\ie
\left\langle\psi^{s_1,s_2}_{P_1}(x_1)\psi^{s_2,s_1}_{P_2}(x_2)\right\rangle=\pi\frac{\delta(P_1-P_2)}{|x_1-x_2|^{2 h_P}},
\fe
where the disc is represented as the upper half plane, with $x_1, x_2 \in\mathbb{R}$. In the weak coupling regime $\phi\to-\infty$, $\psi^{s_1,s_2}_{P}$ admits the free field representation
\ie
\psi_P^{s_1,s_2} \sim (d^{s_1,s_2}(P))^{-{1\over 2}} e^{(1+iP)\phi} + (d^{s_1,s_2}(P))^{{1\over 2}} e^{(1-iP)\phi}.
\fe
where the ``boundary reflection phase" $d^{s_1,s_2}(P)$ is given by \cite{Teschner:2000md}
\ie
d^{s_1,s_2}(P) = \frac{\Gamma_1(2iP)}{\Gamma_1(-2iP)}\frac{S_1(1+i(s_1+s_2-P))S_1(1-i(P+s_1+s_2))}{S_1(1+i(P+s_1-s_2))S_1(1+i(P+s_2-s_1))}.
\fe
Here we have defined the function $S_1(x)=\Gamma_1(x)/\Gamma_1(2-x)$, where $\Gamma_1(x)$ is defined as in section \ref{closedreview}. $S_1(x)$ has poles at $x\in\mathbb{Z}_{\leq0}$ and zeros at $x\in\mathbb{Z}_{\geq2}$, and obeys
\ie
S_1(x&+1)=2\sin(\pi x)S_1(x),\\
S_1(x&)\sim e^{\mp i\frac{\pi}{2}\left(x(x-2)+\frac{5}{6}\right)},~~~~\mathrm{Im}(x) \to \pm\infty.
\fe
The disc 3-point function of boundary primaries takes the form \cite{Ponsot:2001ng, Ponsot:2002ec}
\ie\label{cpppformula}
&\left\langle\psi^{s_1,s_3}_{P_3}(x_3)\psi^{s_3,s_2}_{P_2}(x_2)\psi^{s_2,s_1}_{P_1}(x_1)\right\rangle
\\
& =\frac{C^{s_1,s_2,s_3}(P_1,P_2,P_3)}{|x_3-x_2|^{-h_{P_1}+h_{P_2}+h_{P_3}}|x_1-x_3|^{-h_{P_2}+h_{P_1}+h_{P_3}}|x_2-x_1|^{-h_{P_3}+h_{P_1}+h_{P_2}}},
\fe
where $C^{s_1,s_2,s_3}(P_1,P_2,P_3)$ is the boundary structure constant, given by the formula
\ie
&C^{s_1,s_2,s_3}(P_1,P_2,P_3)=2^{\frac{3}{8}}\pi^{\frac{5}{4}}\frac{\left(d^{s_1,s_3}(P_3)\right)^\frac{1}{2}}{\left(d^{s_2,s_1}(P_1)\right)^\frac{1}{2}\left(d^{s_3,s_2}(P_2)\right)^\frac{1}{2}}\\
& ~~~\times \frac{\Gamma_1(1-i(P_1+P_2+P_3))\Gamma_1(1+i(P_2+P_3-P_1))\Gamma_1(1+i(P_2-P_3-P_1))\Gamma_1(1+i(P_3-P_2-P_1))}{\Gamma_1(2)\Gamma_1(2iP_3)\Gamma_1(-2iP_2)\Gamma_1(-2iP_1)}\\
&~~~\times \frac{S_1(1+i(P_3+s_1-s_3))S_1(1+i(P_3-s_1-s_3))}{S_1(1+i(P_2+s_2-s_3))S_1(1+i(P_2-s_2-s_3))}\int_{\mathbb{R}+i0^+} dt\prod_{k=1}^4\frac{S_1(U_k+i t)}{S_1(V_k+it )}.
\label{eq:bdry3pt}
\fe
In the last line, the variables $U_k$ and $V_k$, $k=1,...,4$, are defined as
\ie
&U_1=1+i(s_1+s_2-P_1),~~~~~~~~U_2=1+i(s_2-s_1-P_1),\\
&U_3=1+i(s_2-s_3+P_2),~~~~~~~~U_4=1+i(s_2-s_3-P_2),\\
&V_1=2+i(s_2-s_3+P_3-P_1),~V_2=2+i(s_2-s_3-P_1-P_3),\\
&V_3=2+2is_2,~~~~~~~~~~~~~~~~~~~~~~V_4=2.\\
\fe
Though not immediately evident from (\ref{cpppformula}), $C^{s_1,s_2,s_3}(P_1,P_2,P_3)$ is real for real $P_1, P_2, P_3$, and is invariant with respect to cyclic permutations of the three boundary primaries. 

Finally, the disc bulk-boundary 2-point function is given by \cite{Hosomichi:2001xc}
\ie
\left\langle V_{P_1}(z,\bar{z})\psi^{s,s}_{P_2}(x) \right\rangle = \frac{\mathcal{R}^{s}(P_1;P_2)}{|z-\bar{z}|^{2h_{P_1}-h_{P_2}}|z-x|^{2h_{P_2}}},
\fe
where $\mathcal{R}^{s}(P_1;P_2)$ is the bulk-boundary structure constant, given by
\ie
\mathcal{R}^{s}(P_1;P_2) &=2^{\frac{5}{8}}\pi^{\frac{3}{4}} (S(P_1))^{-{1\over 2}}\left(d^{s,s}(P_2)\right)^{-{1\over 2}}\frac{\Gamma_1(1-iP_2)^3\Gamma_1(1-i(2P_1+P_2))\Gamma_1(1+i(2P_1-P_2))}{\Gamma_1(2)\Gamma_1(1+iP_2)\Gamma_1(-2iP_2)\Gamma_1(2+2iP_1)\Gamma_1(-2iP_1)}\\
&~~~ \times \int_{-\infty}^{\infty}dt\, e^{4\pi i ts} \frac{S_1(\frac{1}{2}(1+i(2P_1+P_2))+it)S_1(\frac{1}{2}(1+i(2P_1+P_2)-it)}{S_1(\frac{1}{2}(3+i(2P_1-P_2))+it)S_1(\frac{1}{2}(3+i(2P_1-P_2)-it)}.
\label{eq:bulkbdry}
\fe
Though not evident from the above formula, $\mathcal{R}^{s}(P_1;P_2)$ is real for real $P_1, P_2$.


These structure constants are known to obey all consistency conditions based on crossing and modular invariance, and are all that is needed to compute any correlation functions of the $c=25$ Liouville CFT on any Riemann surface with or without boundaries. In particular we have checked numerically the crossing relations for disc bulk 2-point functions and for disc boundary 4-point functions, and verified the consistency of the normalizations of the boundary structure constant and bulk-boundary structure constant, as described in Appendix \ref{sec:appcross} and \ref{sec:appnorm}.

The in and out asymptotic states of an open string stretched between a pair of FZZT branes labeled by $s_1$ and $s_2$ are represented by boundary vertex operators of the form
\ie\label{openstringvop}
\Psi^{s_1,s_2\,\,\pm}_\omega= g_o \bno e^{\pm i \omega X^0}\bno \psi^{s_1,s_2}_{P=\omega},
\fe
where $\bno\cdots \bno$ stands for boundary normal ordering, and the open string coupling $g_o$ is expected to be proportional to $\sqrt{g_s}$. The precise relation between $g_o$ and $g_s$ will be determined in section \ref{sec:WSLLLL}.

A description of FZZT branes in the dual matrix quantum mechanics was suggested in \cite{Gaiotto:2005gd}, but a detailed understanding of the duality map is still missing. On the other hand, a precise duality was conjectured in \cite{Maldacena:2005hi} concerning a limit of open strings on FZZT branes, known as {\it long strings}. The long string is defined as an open string of energy $\omega$ ending on an FZZT brane labeled by the parameter $s$, in the limit
\ie\label{eq:lslimit}
s\to\infty ,~~~\omega\to\infty,\\
\epsilon \equiv \omega - 2s ~~~\mathrm{finite}.
\fe
We will refer to $\epsilon$ as the renormalized energy of the long string. The physical picture is that the FZZT brane recedes to the weak coupling region of space ($\phi \lessapprox - s$), while the open string propagating down the FZZT brane has enough energy to stretch into the bulk and reach the spatial region of finite string coupling, before retracting back to the FZZT brane. To achieve this, the energy of the open string should be comparable to twice that of an open string stretch between the FZZT brane and a ZZ brane (which resides in the strong coupling region), hence $\omega \sim 2s$.


\begin{figure}[h!]
\centering \label{fig:LongString}
\begin{tikzpicture}
\draw [thick] node[above]{FZZT} (-0.5,0.05) -- (1,0.05);
\draw [-,thick, cyan] (1,0.1) to [out=0,in=0] (6,0.05)
to [out=0,in=0] (1,0) ;
\draw[fill] (1,0.05) circle [radius=0.1];
\draw[dashed,ultra thick] (5,-1) to [out=5,in=-100] (9,2.5);
\draw [thick] (-0.5,-1) -- (9,-1) node[right] {$\phi$};
\end{tikzpicture}
\caption{A long string in $c=1$ string theory is the high energy limit of an open string ending on an FZZT brane receding to the weak coupling region.}
\end{figure}
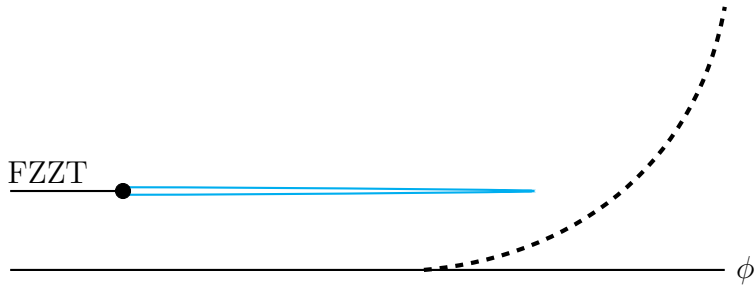

\section{Long string scattering from the worldsheet}
\label{sec:WSamps}

Despite that the long string has infinite energy, most of it lies in the region of space where the string coupling is exponentially suppressed, and one may anticipate well defined scattering amplitudes of the long strings with one another and with closed strings. To formulate the long string scattering amplitudes precisely from the worldsheet, we should keep in mind that the vertex operator (\ref{openstringvop}) corresponds to an open string asymptotic state $|\omega\rangle_o$ that is not quite delta-function normalized, but rather normalized according to ${}_o\langle\omega|\omega'\rangle_o = \omega \delta(\omega-\omega')$ as in the closed string case. Therefore, before taking the long string limit on the open-closed string amplitude,
we need to include an extra factor of $1/\sqrt{\omega}$ associated with each open string vertex operator. That is, the scattering amplitude of long strings ($L$) and closed strings ($C$) of the form $\{L_i, C_j \}\to \{L_k, C_l\}$ is computed as the limit
\ie
\label{eq:lsamplitude}
\cA_{L_iC_j\to L_kC_l}&\left(\{\epsilon_i,\omega_j\}\to\{\epsilon_k,\omega_l\}\right)\equiv\\
&\lim_{s\to\infty}\left(\prod_{i}\frac{1}{\sqrt{\epsilon_i +2s}}\right)\left(\prod_{k}\frac{1}{\sqrt{\epsilon_k +2s}}\right)\cA_{\Psi^{+}_{i}\cV^{+}_{j}\to\Psi^{-}_{k}\cV^{-}_{l}}\left(\{\epsilon_i +2s,\omega_j\}\to\{\epsilon_k +2s,\omega_l\}\right),
\fe
where $\cA_{\Psi^{+}_{i}\cV^{+}_{j}\to\Psi^{-}_{k}\cV^{-}_{l}}$ denotes the amplitude computed using the open string vertex operators (\ref{openstringvop}) and closed string vertex operators (\ref{ima}). Here we have retained the normalization (\ref{closednormal}) for closed string states, for convenience in comparison with matrix model results \cite{Balthazar:2017mxh}. Note that we could have replaced the factor $1/\sqrt{\epsilon+2s}$ on the RHS of (\ref{eq:lsamplitude}) by simply $1/\sqrt{2s}$ in the limit $s\to \infty$, but the former is more convenient for numerical extrapolation to $s=\infty$ as the resulting amplitude converges exponentially fast in $s$.

\subsection{The long $\to$ long $+$ closed string amplitude}

We begin with the tree level open $\to$ open $+$ closed string amplitude $\cA_{\Psi^{s,s\,+}_{\omega_1}\to\Psi^{s,s\,-}_{\omega_2}\cV^{-}_{\omega_3}}$, computed as the disc bulk-boundary 3-point function integrated with respect to one modulus parameterizing the location of one of the open string vertex operators,
\ie
&2\int_0^\infty dx\, \left\langle \Psi^{s,s\,\,+}_{\omega_1}(0) \Psi^{s,s\,\,-}_{\omega_2}(x) \cV^{-}_{\omega_3}(i/2) \right\rangle = 2i g_o^2 g_s C_{D^2}\delta(\omega_1-\omega_2-\omega_3) \\
&~~~~~~~~~~~\times \int_0^\infty dx\, {1\over 4}  \, 2^{-2\omega_1\omega_3}|x|^{2\omega_1\omega_2}|x-i/2|^{-2\omega_2\omega_3} \left\langle \psi^{s,s}_{\omega_1}(0) \psi^{s,s}_{\omega_2}(x) V_{\omega_3/2}(i/2) \right\rangle_{\text{Liouville}}.
\label{eq:LLCfull}
\fe
On the RHS, $C_{D^2}$ is a normalization constant associated with the disc topology. 
The factor ${1\over 4}$ in the integrand is due to the ghost correlator. $\langle\cdots\rangle_{\rm Liouville}$ stands for the disc correlator in the $c=25$ Liouville theory subject to FZZT boundary condition. It can be 
computed as an integral of the boundary Liouville structure constants multiplied by the relevant Virasoro conformal block,
\ie\label{eq:LLCLiouv}
& \langle \psi^{s,s}_{\omega_1}(0)  \psi^{s,s}_{\omega_2}(x) V_{\omega_3/2}(i/2) \rangle_{\text{Liouville}} = \,\, 2^{2 h_1} |x-i/2|^{-2h_2}\left(\frac{x-i/2}{x+i/2}\right)^{-h_1} \\
&~~~~~~~~~~~ \times\int_0^{\infty}\frac{dP}{\pi}\mathcal{R}^{s}(\omega_3/2;P)C^{s,s,s}(P,\omega_1,\omega_2)\,i^{-h_1+h_2-h}F(h_1,h_2,h_3,h_3;h\left|\eta\right.).
\fe
Here ${\cal R}^s$ is the boundary-bulk structure constant, and $C^{s,s,s}$ is the boundary structure constant, as introduced in section \ref{fzztbraneintro}. The disc boundary-bulk 3-point conformal block in question is equivalent to a sphere 4-point holomorphic Virasoro conformal block $F(h_1,h_2,h_3,h_3;h\left|\eta\right.)$ with $c=25$, external weights $h_1=1+\omega_1^2,\,h_2=1+\omega_2^2,\,h_3=1+\omega_3^2/4$, internal weight $h=1+P^2$, evaluated at the cross-ratio $\eta=2x/(x-i/2)$. We follow the conventions of Appendix C of \cite{Chang:2014jta} for the Virasoro conformal block, and the extra phase factor in (\ref{eq:LLCLiouv}) is needed to ensure the reality of the correlation function.

The moduli integral in (\ref{eq:LLCfull}) is a priori divergent near $x=0$, where the two boundary vertex operators collide, and must be regularized. Such divergences in the moduli integration are familiar in string perturbation theory, and is usually regularized by suitable analytic continuation in the external momenta. In our case, such an analytic continuation is inaccessible as we would like to evaluate the amplitude (\ref{eq:LLCfull}) numerically at physical energies and compare directly with results in the dual matrix model. To proceed, we adopt the regularization method introduced in \cite{Balthazar:2017mxh} by subtracting suitable counter terms from the moduli integrand. More explicitly, near $x=0$, the moduli integrand takes the form 
\ie\label{intgrandxx}
\int \frac{dP}{\pi}\mathcal{R}^{s}(\omega_3/2;P)C^{s,s,s}(P,\omega_1,\omega_2)2^{2+2P^2-2\omega_3^2}x^{-1+P^2-\omega_3^2}\sum_{n=0}a_n x^n,
\fe
where $a_0=1$, and the remaining coefficients $a_{n>0}$ can be obtained from expanding the Virasoro conformal block as well as the prefactors appearing in (\ref{eq:LLCfull}). The $P$-integration in the range $P\leq\omega_3$ leads to singular terms in $x$ at $x=0$. This is regularized by subtracting from (\ref{intgrandxx}) the counter term
\ie
R(x) = \sum_{0\leq n\leq \omega_3}a_n \int_0^{\sqrt{\omega_3^2-n}}\frac{dP}{\pi}\mathcal{R}^{s}(\omega_3/2;P)C^{s,s,s}(P,\omega_1,\omega_2)2^{2+2P^2-2\omega_3^2}x^{-1+P^2-\omega_3^2+n}.
\fe
There is a remaining potential divergence due to the contribution from the $P$-integral in a small neighborhood of $P=\omega_3$, which can be dealt with by assigning a small imaginary part to $\omega_3$ and $\epsilon_1$. With this understanding, the regularized moduli integral is given by
\ie
\int_0^\infty dx \left[2^{-2\omega_1\omega_3}|x|^{2\omega_1\omega_2}|x-i/2|^{-2\omega_2\omega_3} \langle \psi^{s,s}_{\omega_1}(0) \psi^{s,s}_{\omega_2}(x) V_{\omega_3/2}(i/2) \rangle_{D^2,\text{Liouville}} - R(x) \vphantom{\frac{1}{1}}\right]
\label{eq:LLCreg}.
\fe
Note that the integrand behaves as $x^{-2}$ for large $x$.

A posteriori, numerical computations indicate that in the long string limit, the amplitude (\ref{eq:LLCfull}) is in fact dominated by the $P$-integral over an intermediate range of momenta $P\sim \omega_1$, which is far from the region where the above prescribed regularization is needed. Consequently, to compute the amplitude in the long string limit, one can simply perform the $P$-integration starting from $P$ slightly above $\omega_3$ to avoid the divergence near $x=0$, and it suffices to work directly with real $\omega_3$ and $\epsilon_1$. More details are given in appendix \ref{sec:appnint}.


\subsubsection{Numerical results}


The Virasoro conformal block appearing in (\ref{eq:LLCLiouv}) can be evaluated efficiently using Zamolodchikov's recurrence relations \cite{Zamolodchikov:1985ie} as an expansion in the elliptic nome $q$, related to the cross ratio $\eta$ by $q=\exp[ -\pi K(1-\eta)/K(\eta)]$, where $K(\eta)={}_2F_1(1/2,1/2,1;\eta)$.
The $q$-expansion converges on the entire complex $\eta$-plane away from $\eta=1$ and $\infty$. For our application, sufficiently accurate numerical results can be obtained simply by truncating the $q$-expansion at low orders. We then multiply the conformal block by the structure constants, and numerically integrate over the intermediate weight/Liouville-momentum, and over the modulus $x$. Further details of the numerics are described in appendix \ref{sec:appnint}. 

\begin{figure}[h!]
\centering
\includegraphics[width=10cm]{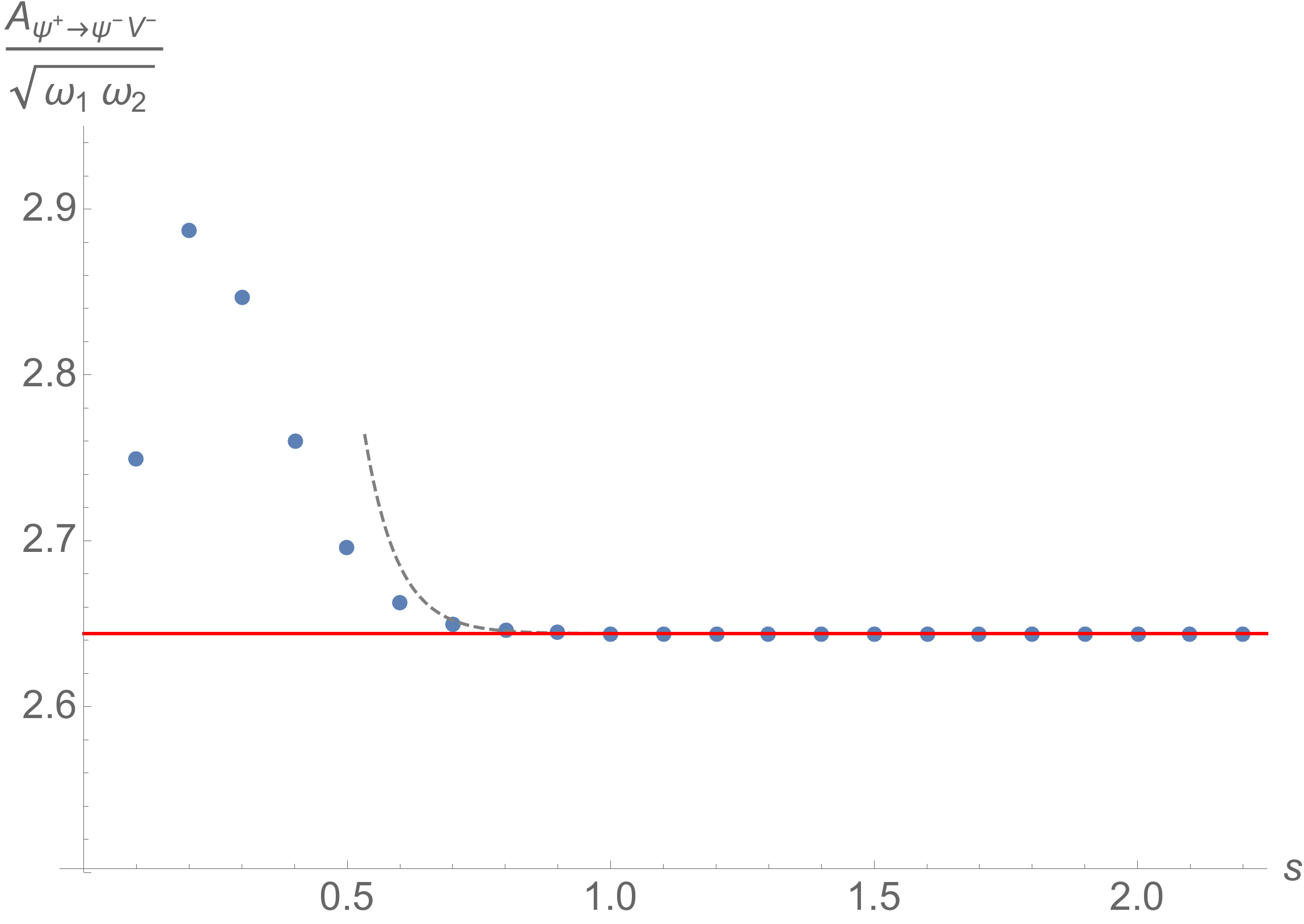}
\caption{A sample of the amplitude $\cA_{\Psi^{s,s\,+}_{\omega_1}\to\Psi^{s,s\,-}_{\omega_2}\cV^{-}_{\omega_3}}/\sqrt{\omega_1 \omega_2}$, dropping the prefactors in the first line of (\ref{eq:LLCfull}), 
evaluated at $\epsilon_2=0.7$, $\omega_3=0.5$, at increasing $s$, is shown in blue dots. The dashed grey line is an exponential fit of the amplitudes at large $s$, whereas the solid red line marks the long string limit of the amplitude.}
\label{fig:LLClslimit}
\end{figure}

We proceed by evaluating $\cA_{\Psi^{s,s\,+}_{\omega_1}\to\Psi^{s,s\,-}_{\omega_2}\cV^{-}_{\omega_3}}$ numerically at fixed renormalized long string energies $\epsilon_1, \epsilon_2$, with increasing $s$. According to (\ref{eq:lsamplitude}), we must also include an overall factor of $1/\sqrt{\omega_1\omega_2}$, and then take the long string limit $s\to \infty$. In this limit, the FZZT brane recedes to the asymptotic region where the effective string coupling is suppressed exponentially. Therefore, we expect exponentially fast convergence of the amplitude ${1\over \sqrt{\omega_1 \omega_2}}\cA_{\Psi^{s,s\,+}_{\omega_1}\to\Psi^{s,s\,-}_{\omega_2}\cV^{-}_{\omega_3}}$ in the long string limit. This is indeed confirmed by numerical results, as in the example shown in Figure \ref{fig:LLClslimit}.

\begin{figure}[h!]
\centering
{\includegraphics[width=0.8\textwidth]{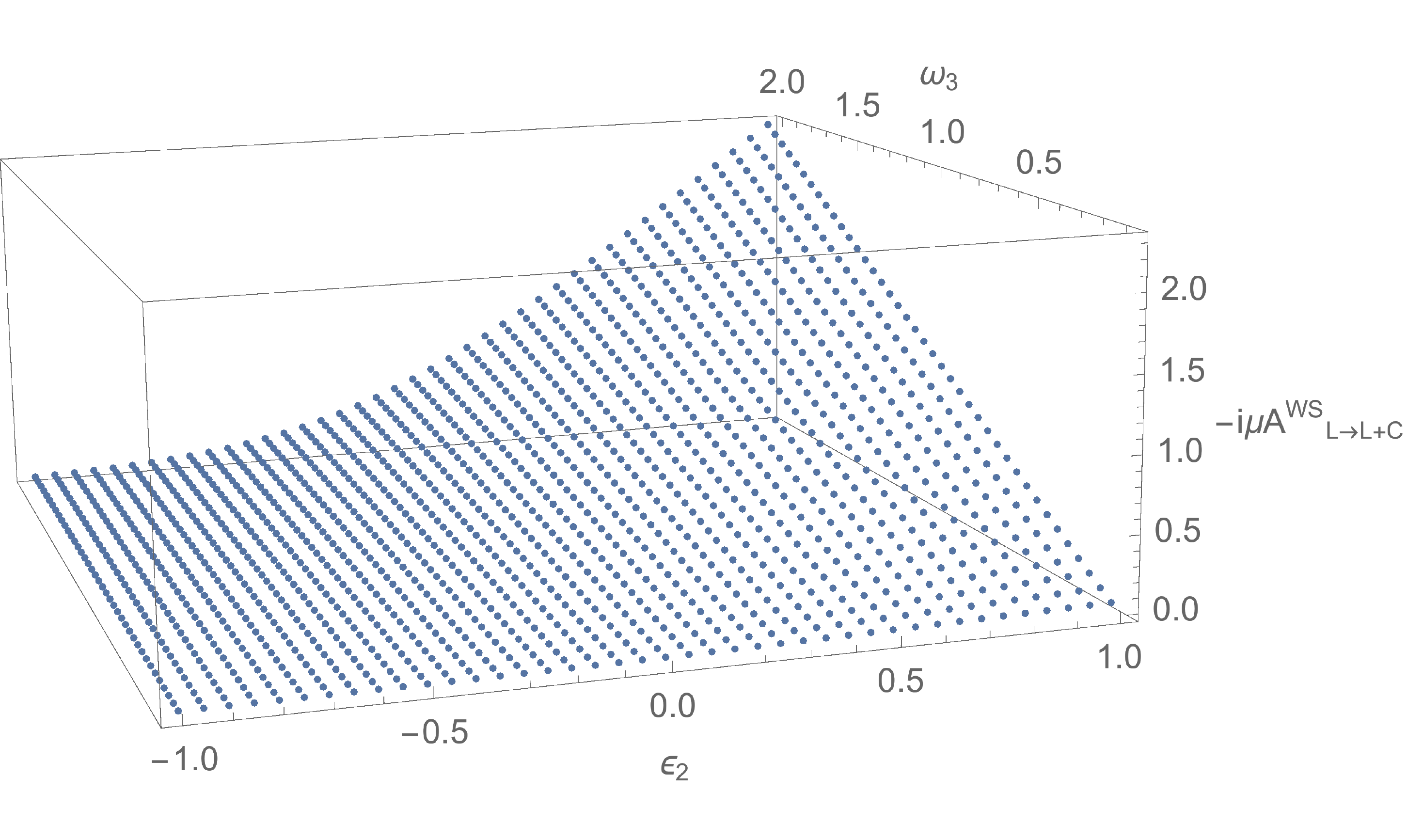}}
\caption{The long $\to$ long + closed string amplitude $\cA_{L\to L+C}$ (rescaled by $\mu$) as a function of the outgoing long string energy $\epsilon_2$ and the closed string energy $\omega_3$. 
}
\label{fig:LLCampdep}
\end{figure}

\begin{figure}[h!]
\centering
\subfloat[][$\omega_3=1$.]{\includegraphics[width=0.45\textwidth]{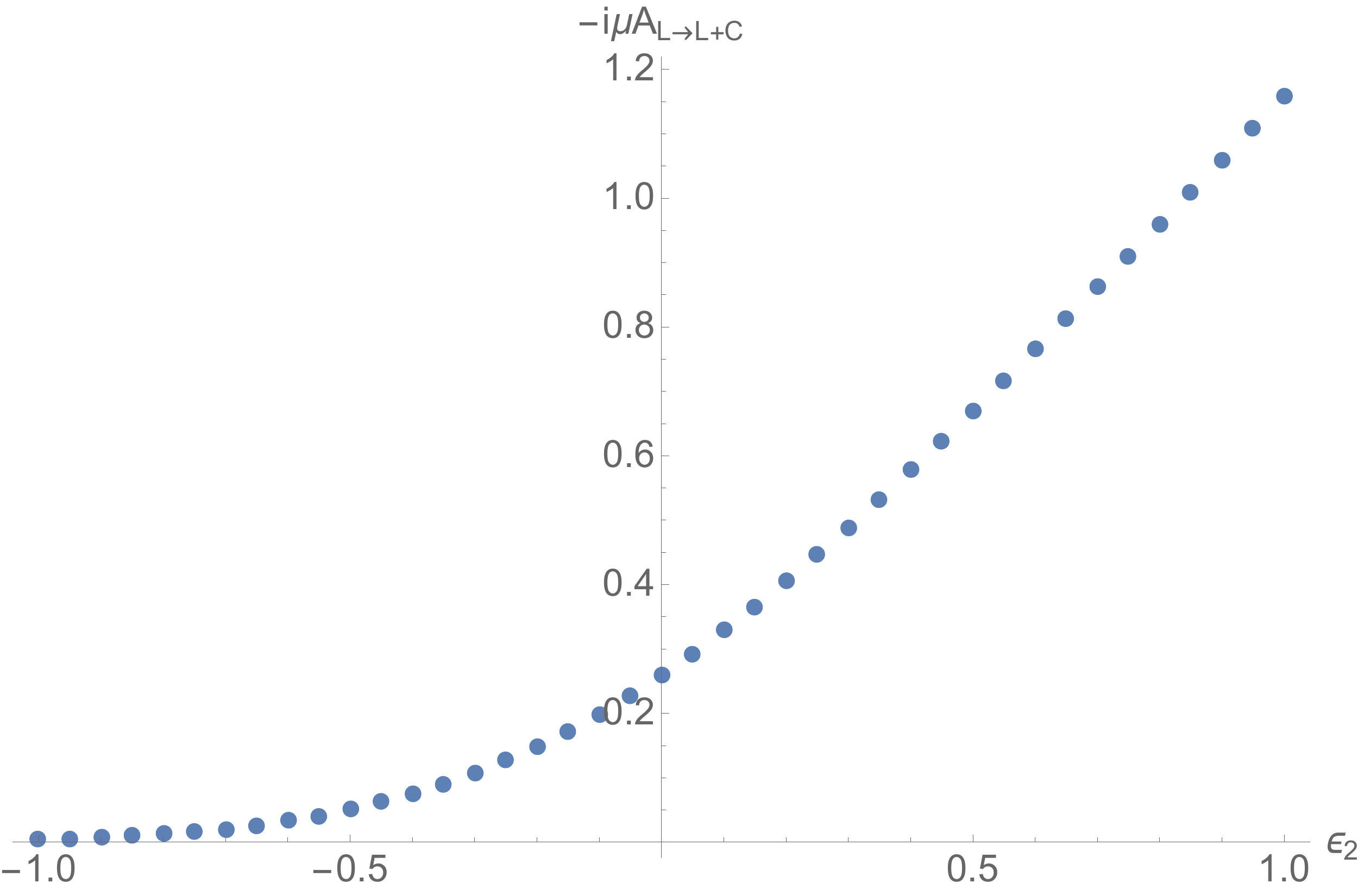}\label{fig:LLCe2depfixw}}~
\subfloat[][$\omega_3=2.5$.]{\includegraphics[width=0.45\textwidth]{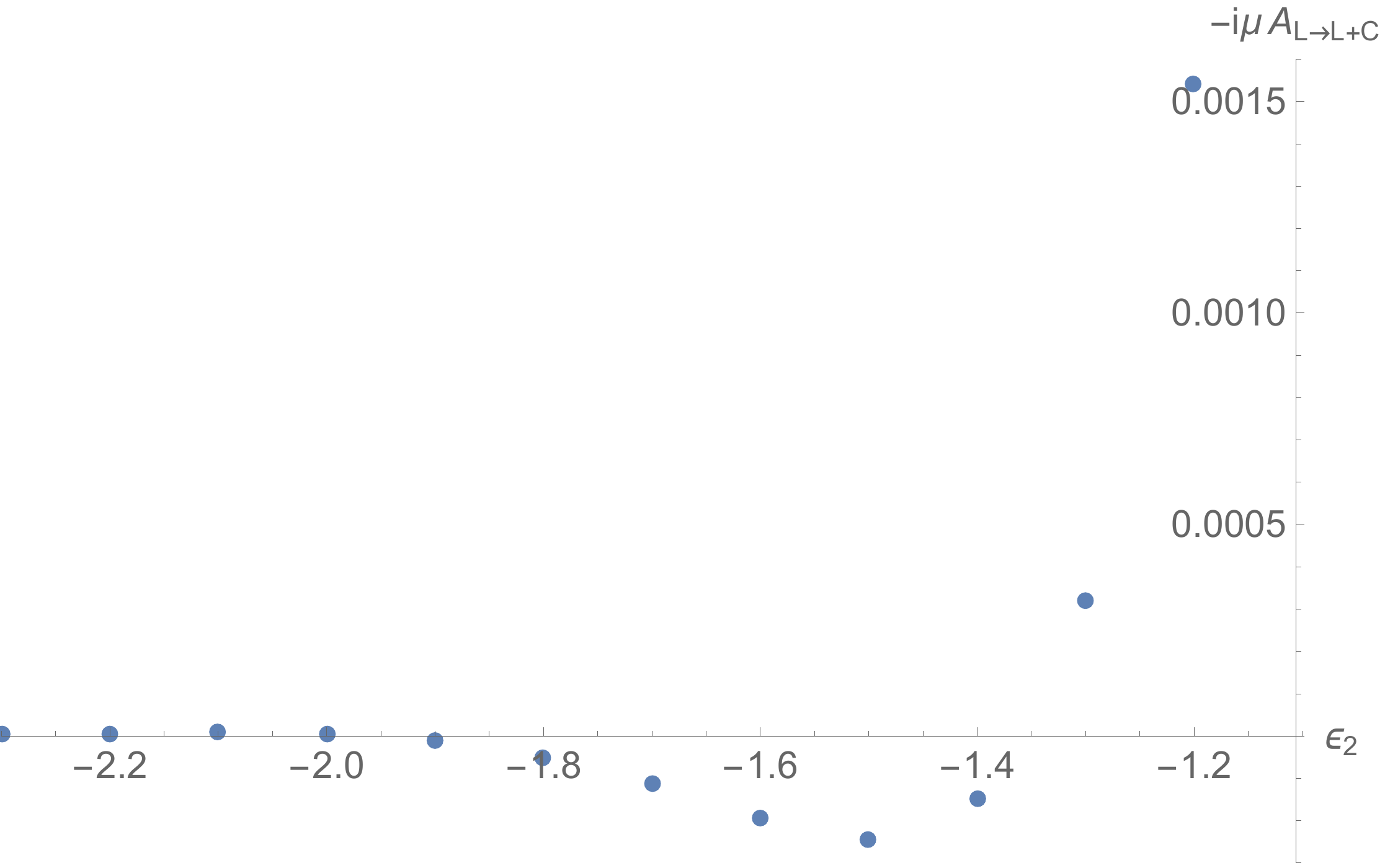}\label{fig:LLCnege2dep}}\\
\subfloat[][$\omega_3=2.5$.]{\includegraphics[width=0.45\textwidth]{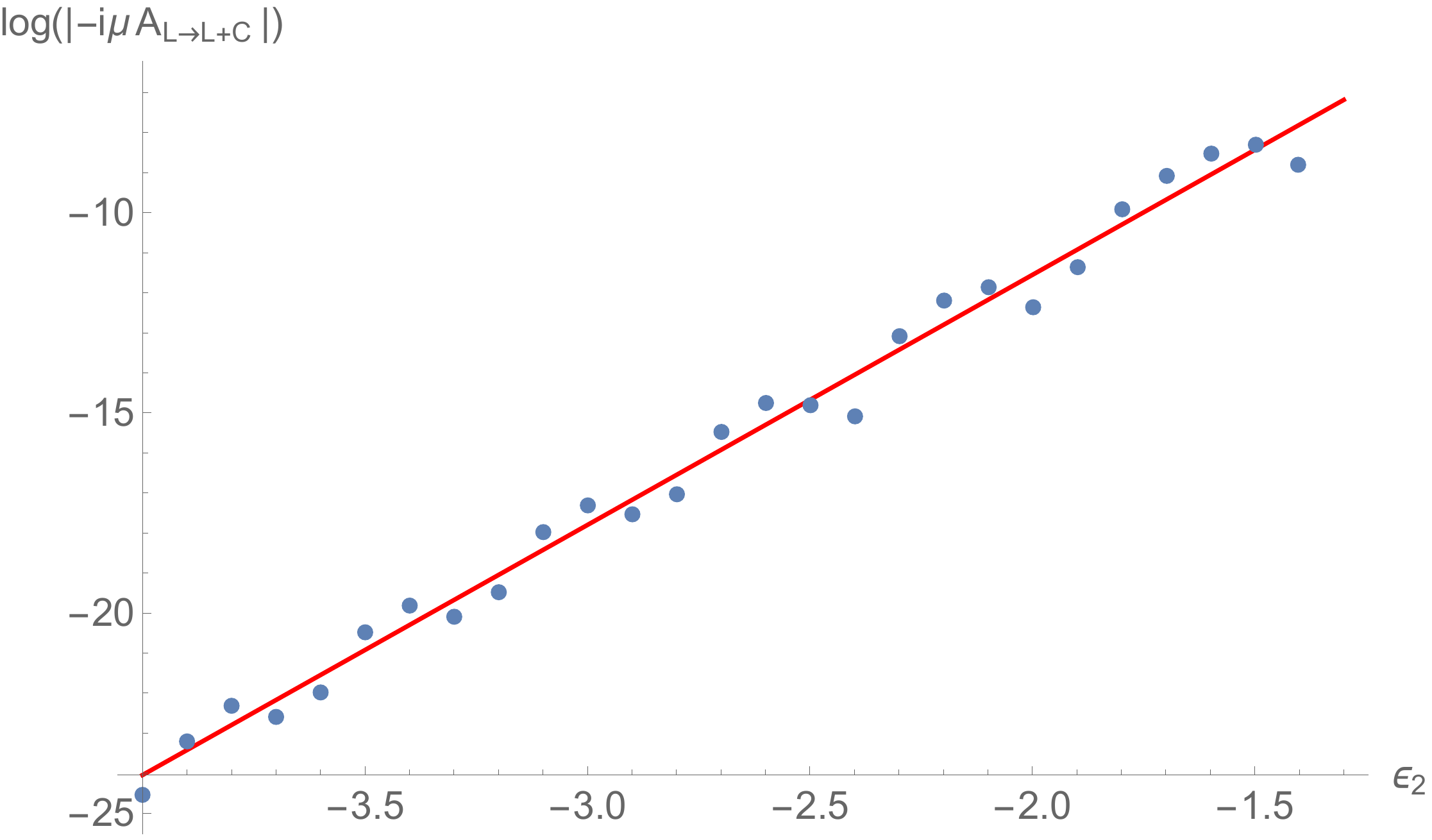}\label{fig:LLCexpfit}}
\caption{Long string amplitude $\cA_{L\to L+C}$ evaluated at fixed $\omega_3$ as a function of $\epsilon_2$. Here we have 
rescaled the amplitude by $\mu={1\over 2\pi g_s}$. In (c) we plot the logarithm of the absolute value of the amplitude over a range of sufficiently large negative $\epsilon_2$. The red line represents a linear fit of slope $6.25$, which is in reasonable agreement with our expectation that the amplitude is modulated by a $e^{2\pi\epsilon_2}$ profile in this regime.
}
\label{fig:LLCfixedw3}
\end{figure}

\begin{figure}[h!]
\centering
\subfloat[][$\epsilon_2=0.5$.]{\includegraphics[width=0.45\textwidth]{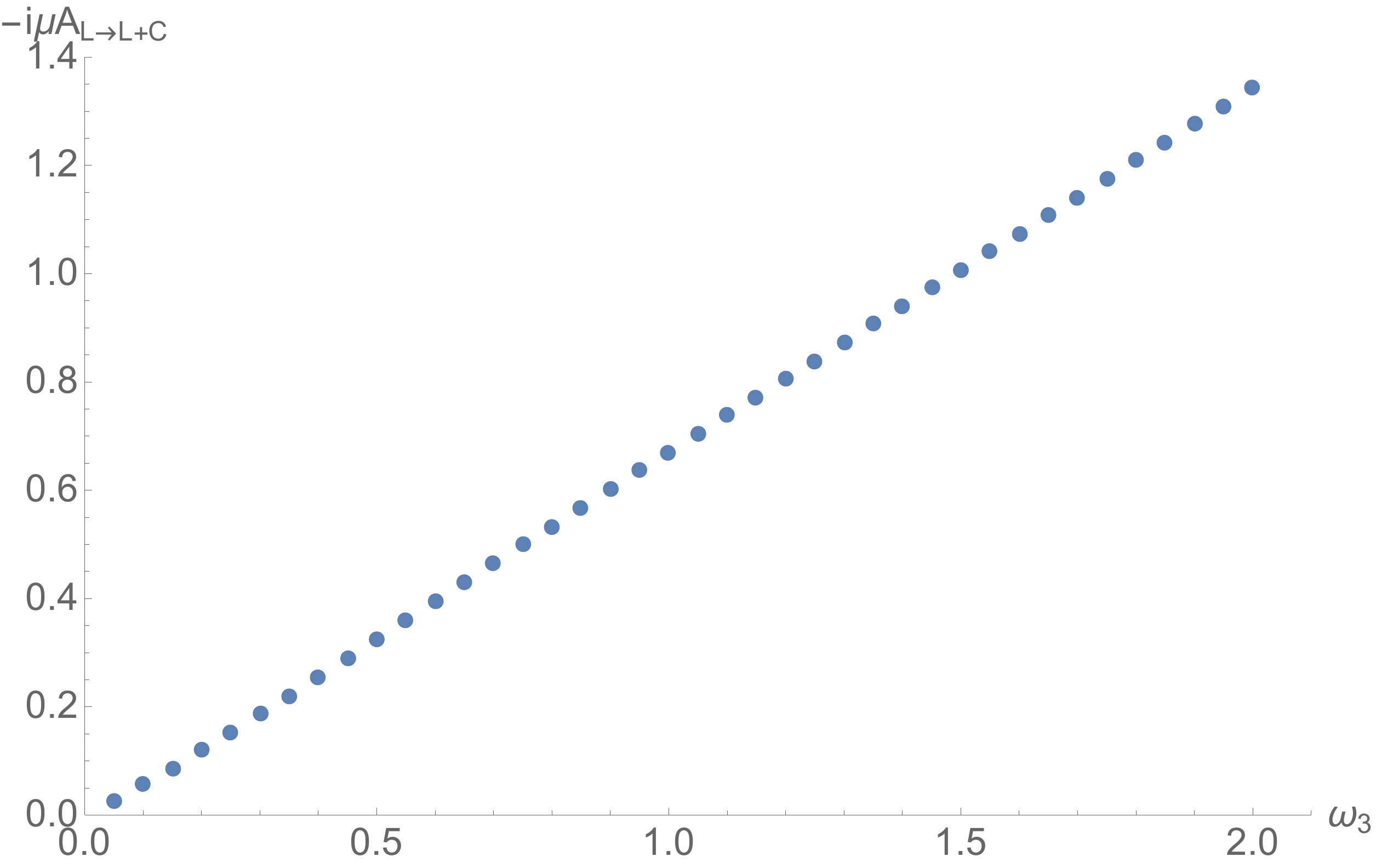}\label{fig:LLCw3dep}}\\
\subfloat[][$\epsilon_2=-2.0$.]{\includegraphics[width=0.45\textwidth]{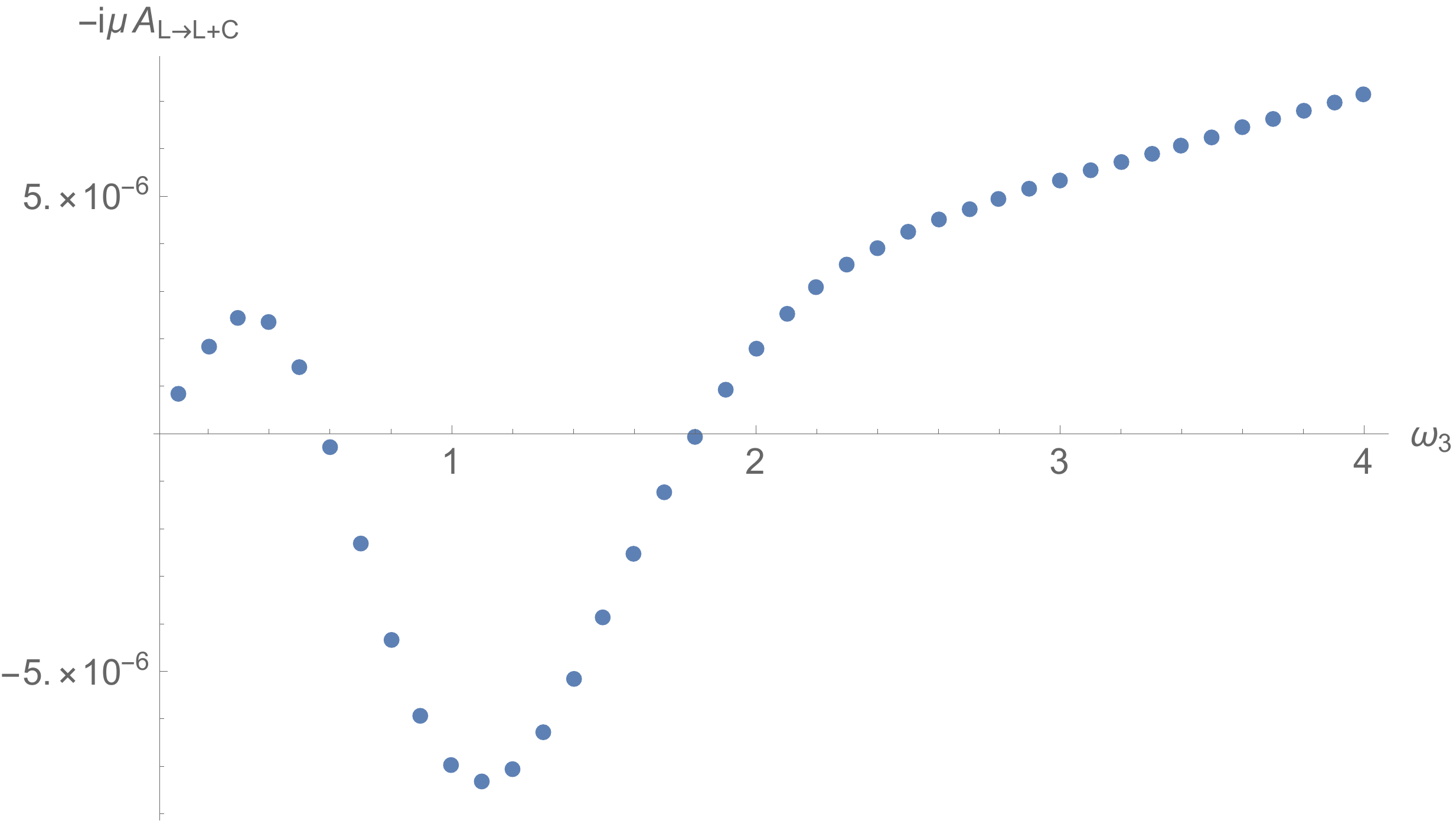}\label{fig:LLCe2m2}}~
\subfloat[][$\epsilon_2=-3.0$.]{\includegraphics[width=0.45\textwidth]{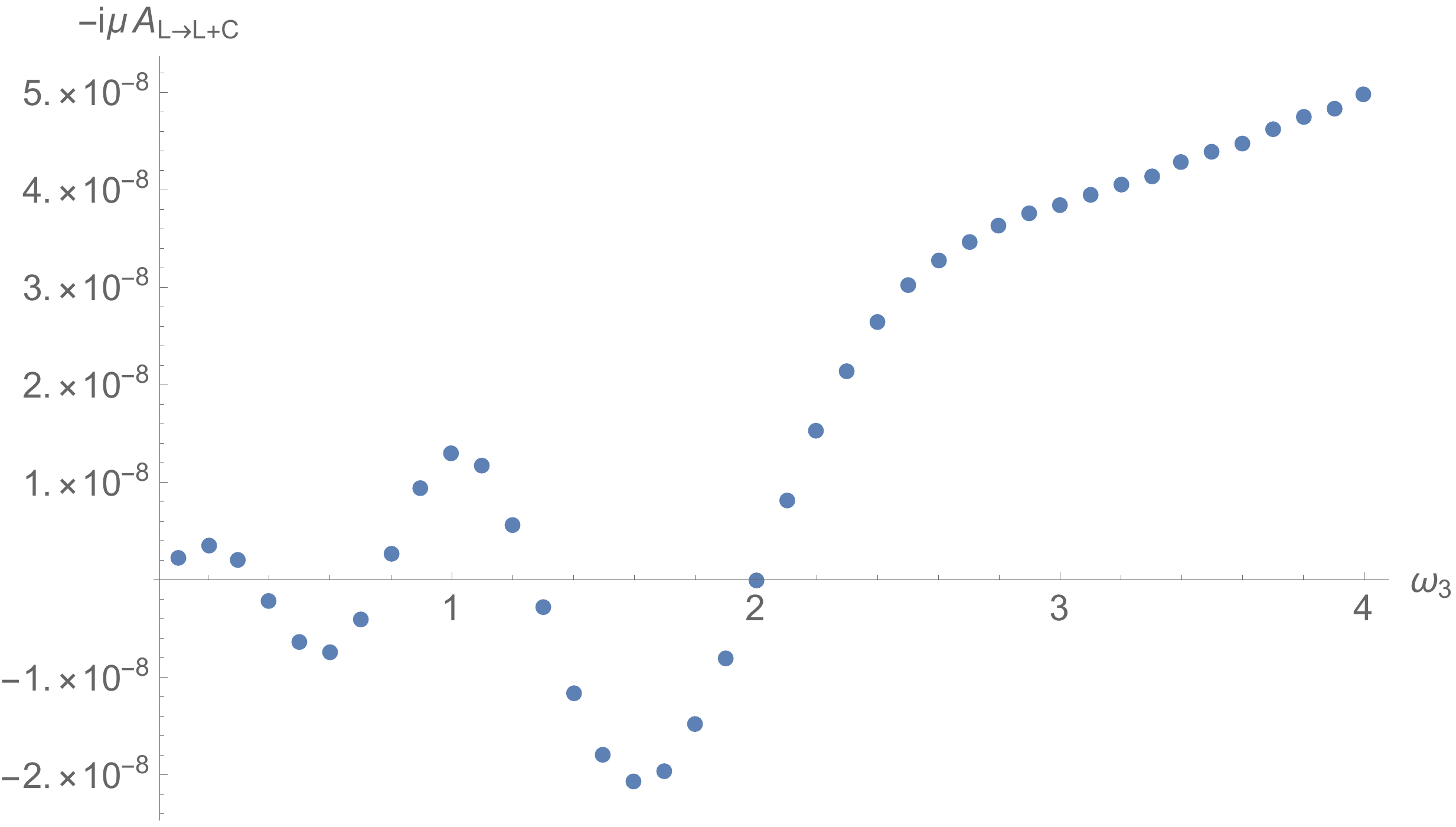}\label{fig:LLCe2m3}}
\caption{Long string amplitude $\cA_{L\to L+C}$ (rescaled by $\mu$) evaluated at fixed outgoing long string energy $\epsilon_2$ as a function of closed string energy $\omega_3$.
}
\label{fig:LLCfixede2}
\end{figure}

\begin{figure}[h!]
\centering
{\includegraphics[width=0.45\textwidth]{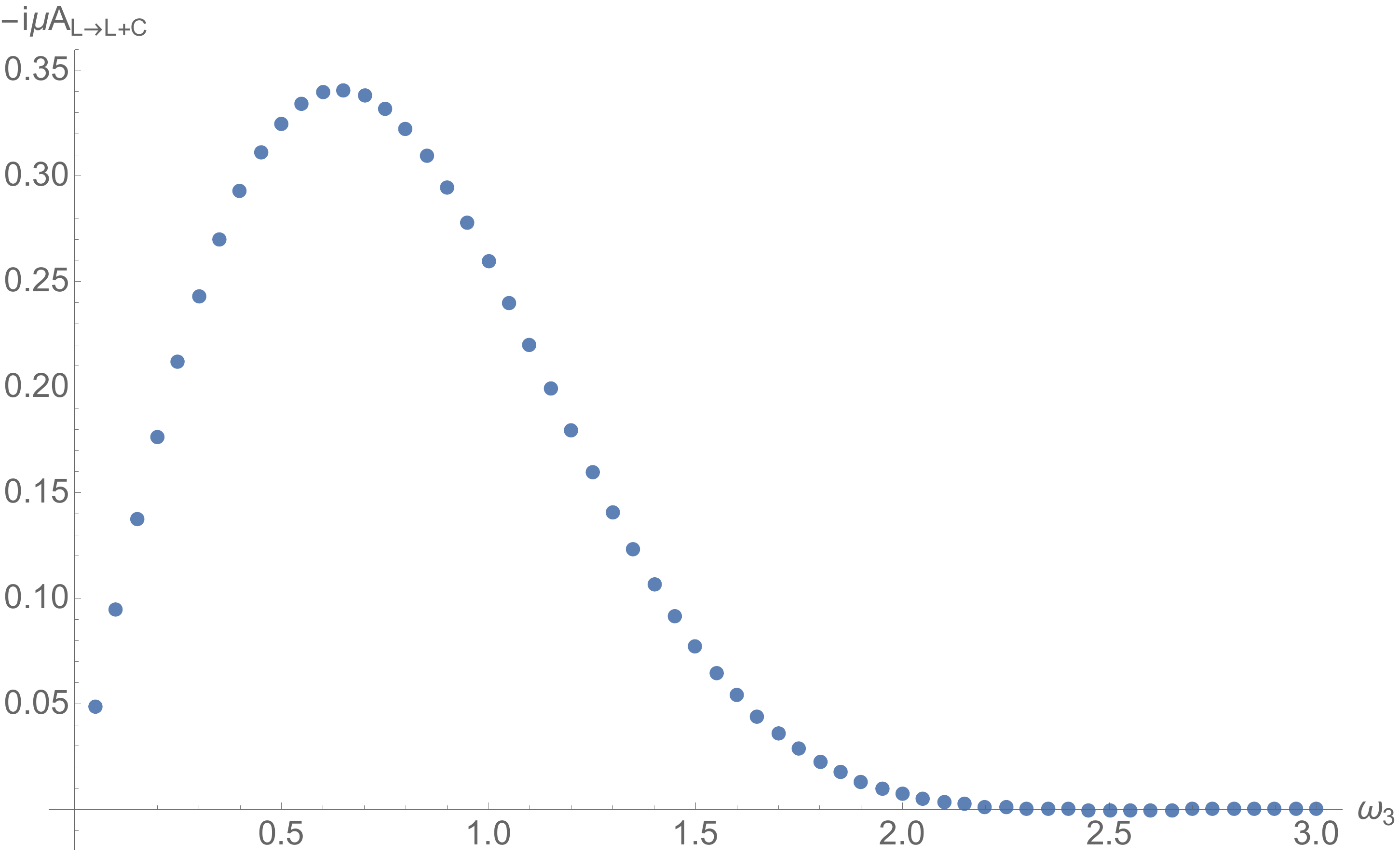}}
\caption{Long string amplitude $\cA_{L\to L+C}$ (rescaled by $\mu$) evaluated at fixed incoming long string energy $\epsilon_1=1$ as a function of closed string energy $\omega_3$. 
}
\label{fig:LLCfixede1}
\end{figure}

The long $\to$ long + closed amplitude $\cA_{L\to L+C}$, defined as the limit in (\ref{eq:lsamplitude}), is then obtained from an exponential numerical fit. We perform this calculation over a range of renormalized energy $\epsilon_2$ of the outgoing long string, and over a range of the closed string energy $\omega_3$, and plot the results in Figures \ref{fig:LLCampdep}, \ref{fig:LLCfixedw3}, \ref{fig:LLCfixede2}, and \ref{fig:LLCfixede1}. 

Recall that from the closed string amplitude computation in \cite{Balthazar:2017mxh} that sphere amplitudes come with a normalization factor  $C_{S^2}=2\pi/g_s^2$, and $g_s$ is related to the parameter $\mu$ in the dual matrix model via $2\pi g_s=1/\mu$. In principle, consideration of unitarity for open string amplitudes on FZZT branes (before taking the long string limit) would allow for fixing $C_{D^2}$ in terms of $g_o$. In this paper, we will simply determine $C_{D^2}$ by comparing ${\cal A}_{L\to L+C}$ to the matrix model amplitude (\ref{eq:LLCampMM}) computed in the next section. Our numerical results suggest the identification
\ie
\label{eq:CD2}
C_{D^2} = \frac{1}{2g_o^2}
\fe
to high accuracy.

Figure \ref{fig:LLCfixedw3} shows the amplitude $\cA_{L\to L+C}$ at a fixed value of the closed string energy $\omega_3$, as a function of the renormalized energy of the outgoing long string $\epsilon_2$. Its qualitative features can be partially understood as follows. For $\epsilon_2\ll -1$, the tip of the long string always remains in the weak coupling region, and reaches at most $\phi \sim \epsilon_2/(2T)$, where $T={1\over 2\pi}$ is the tension of the string. At the latter location, the effective string coupling is $g_s e^{2\phi} \sim g_s e^{2\pi\epsilon_2}$, indicating that the amplitude of emitting a closed string is exponentially suppressed. This is consistent with our numerical results, in that the latter exhibits an oscillatory behavior at sufficiently large negative $\epsilon_2$ modulated by an exponential suppression profile that fits with $\sim e^{2\pi \epsilon_2}$, as shown in Figure \ref{fig:LLCexpfit}. This exponential decay profile can also be seen by considering the amplitude $\cA_{L\to L+C}$ for fixed incoming renormalized energy $\epsilon_1$ and increasing outgoing closed string energy $\omega_3$, as shown in Figure \ref{fig:LLCfixede1}.

On the other hand, for $\epsilon_2\gg 1$, we expect the tip of the long string to reach deep into the Liouville barrier, up to the location $\phi \sim {1\over 2} \ln \epsilon_2$ where the renormalized energy of the long string is dominated by the Liouville potential energy $\sim e^{2\phi}$. Now the effective string coupling at the tip of the long string is $g_s e^{2\phi} \sim g_s \epsilon_2$, giving rise to the linear behavior of the amplitude in $\epsilon_2$, as seen in Figure \ref{fig:LLCe2depfixw} for sufficiently large energy $\epsilon_2$.

Figure \ref{fig:LLCfixede2} shows $\cA_{L\to L+C}$ as a function of the closed string energy $\omega_3$ (which is positive by definition), at a fixed value of the outgoing long string renormalized energy $\epsilon_2$. The result is approximately linear for sufficiently large $\omega_3$, and oscillatory for small $\omega_3$. We do not know a semi-classical explanation of this behavior.

\subsubsection{A resonance computation}

While we do not have a closed form analytic expression for ${\cal A}_{L\to L+C}$, it is possible to consider the analytic continuation of the amplitude as a function of $\omega_3$ to a special ``resonance energy", analogously to the resonance computation for the closed string amplitudes \cite{DiFrancesco:1991ocm,DiFrancesco:1991daf,Balthazar:2017mxh}.
In particular, if we analytically continue the open $\to$ open + closed string amplitude on a FZZT brane, $\cA_{\Psi^{s,s\,+}_{\omega_2+i}\to\Psi^{s,s\,-}_{\omega_2}\cV^{-}_{i}}$, to imaginary closed string energy $\omega_3=i$ while fixing the open string energy $\omega_2$, the integration over the intermediate Liouville momentum $P$ in (\ref{eq:LLCLiouv}) becomes dominated by the contribution near $P=0$, where the structure constants also simplify. This allow us to obtain an exact answer for the amplitude at $\omega_3=i$,
\ie
\cA_{\Psi^{s,s\,+}_{\omega_2+i}\to\Psi^{s,s\,-}_{\omega_2}\cV^{-}_{i}}=\frac{\pi}{4} g_s \frac{\sinh(2\pi\omega_2)}{\sinh(\pi\omega_2)}\left[\frac{(2i\omega_2)(-1+2i\omega_2)^2(-2+2i\omega_2)}{\sinh(\pi\omega_2+2\pi s)\sinh(\pi\omega_2-2\pi s)}\right]^{\frac{1}{2}}
\label{eq:LLCres}
\fe
Details of the calculation are presented in appendix \ref{sec:appres}.

(\ref{eq:LLCres}) exhibits some expected qualitative behavior, such as exponential decay for $\epsilon_2 \equiv \omega_2-2s\ll 0$, and the absence of exponential growth for $\epsilon_2\gg 0$. However, it diverges at $\epsilon_2=0$. In the limit of large $s$ with fixed $\epsilon_2$, (\ref{eq:LLCres}) diverges as $\sim s^2$. The corresponding amplitude with normalized open string asymptotic states would diverge linearly in $s$. Thus, unlike the case of real energies, the amplitude analytically continued to $\omega_3=i$ does not have a well defined long string limit. We do not have a semi-classical explanation of these divergences.

\subsection{The long $+$ long $\to$ long $+$ long string amplitude}
\label{sec:WSLLLL}

\begin{figure}[h]
\centering
\subfloat[]
{~~~~~
\begin{tikzpicture}
\filldraw[color=black, fill=black!10, thick] (0,0) circle (1);
\draw (1,0) node[cross=4pt, very thick] {};
\draw (-1,0) node[cross=4pt, very thick] {};
\draw (0,-1) node[cross=4pt, very thick] {};
\draw (0,1) node[cross=4pt, very thick] {};
\draw (1,0) node[right] {$\psi_{P_3}$};
\draw (-1,0) node[left] {$\psi_{P_1}$};
\draw (0,-1) node[below] {$\psi_{P_2}$};
\draw (0,1) node[above] {$\psi_{P_4}$};

 \draw (-0.866025,-0.5) arc (-75:75:0.51);
 
  \draw[dashed] (0.866025,0.5) arc (105:255:0.51);
  
  \draw (0.5,-0.866025) arc (15:165:0.51);
  
  \draw[dashed] (-0.5,0.866025) arc (195:345:0.51);

\end{tikzpicture}
}~~
\subfloat[]
{~~~~~
\begin{tikzpicture}
\filldraw[color=black, fill=black!10, thick] (0,0) circle (1);
\draw (1,0) node[cross=4pt, very thick] {};
\draw (-1,0) node[cross=4pt, very thick] {};
\draw (0,-1) node[cross=4pt, very thick] {};
\draw (0,1) node[cross=4pt, very thick] {};
\draw (1,0) node[right] {$\psi_{P_2}$};
\draw (-1,0) node[left] {$\psi_{P_1}$};
\draw (0,-1) node[below] {$\psi_{P_3}$};
\draw (0,1) node[above] {$\psi_{P_4}$};

 \draw (-0.866025,-0.5) arc (-75:75:0.51);
 
  \draw (0.866025,0.5) arc (105:255:0.51);
  
  \draw[dashed] (0.5,-0.866025) arc (15:165:0.51);
  
  \draw[dashed] (-0.5,0.866025) arc (195:345:0.51);

\end{tikzpicture}
}~~
\subfloat[]
{~~~~~
\begin{tikzpicture}
\filldraw[color=black, fill=black!10, thick] (0,0) circle (1);
\draw (1,0) node[cross=4pt, very thick] {};
\draw (-1,0) node[cross=4pt, very thick] {};
\draw (0,-1) node[cross=4pt, very thick] {};
\draw (0,1) node[cross=4pt, very thick] {};
\draw (1,0) node[right] {$\psi_{P_4}$};
\draw (-1,0) node[left] {$\psi_{P_1}$};
\draw (0,-1) node[below] {$\psi_{P_3}$};
\draw (0,1) node[above] {$\psi_{P_2}$};

 \draw (-0.866025,-0.5) arc (-75:75:0.51);
 
  \draw[dashed] (0.866025,0.5) arc (105:255:0.51);
  
  \draw[dashed] (0.5,-0.866025) arc (15:165:0.51);
  
  \draw (-0.5,0.866025) arc (195:345:0.51);

\end{tikzpicture}
}~~
\caption{Three of the Liouville disc diagrams that contribute to long $+$ long $\to$ long $+$ long string scattering. The other three diagrams (not shown) are related by exchanging $P_3\leftrightarrow P_4$. The solid and dashed lines represent incoming and outgoing long strings respectively. Diagrams (a) and (c) are suppressed as the interaction between the incoming strings occurs on the boundary.}
\label{fig:LLLL}
\end{figure}
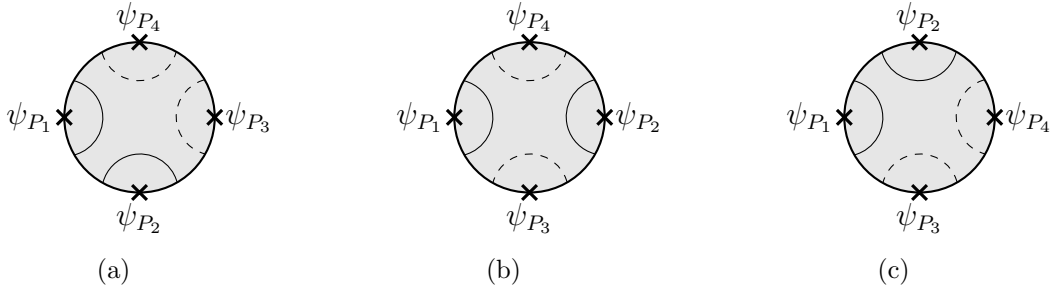

Now we turn to the scattering of a pair of long strings. At tree level, this is given by the long string limit of an open string disc 4-point amplitude
\ie\label{eq:LLLLfull}
&\cA_{\Psi^{s,s\,\,+}_{\omega_1}\Psi^{s,s\,\,+}_{\omega_2}\to\Psi^{s,s\,\,-}_{\omega_3}\Psi^{s,s\,\,-}_{\omega_4}}= i g_o^4C_{D^2}\delta(\omega_1+\omega_2-\omega_3-\omega_4)\\
&~~~~~~~~\times\sum_{j=1}^{3}\int_{I_j}dx\, |x|^{-2\omega_1\omega_2}|x-1|^{2\omega_2\omega_3} \langle \psi^{s,s}_{\omega_1}(0)\psi^{s,s}_{\omega_2}(x)\psi^{s,s}_{\omega_3}(1)\psi'^{s,s}_{\omega_4}(\infty) \rangle_{\text{Liouville}}+(3\leftrightarrow 4),
\fe
where $I_j$ ($j=1,2,3$) denotes integration domains corresponding to the three different configurations of the disc diagram, as shown in Figure \ref{fig:LLLL}. 
Here we have taken the open strings to end on a single FZZT brane labeled by the parameter $s$, as opposed to different FZZT branes. This difference will be unimportant in the long string limit. 

In each integration domain $I_j$, the Liouville disc boundary 4-point function appearing on the RHS of (\ref{eq:LLLLfull}) will be computed by integrating the disc Virasoro conformal block multiplied by the appropriate boundary structure constants, over an internal weight or Liouville momentum $P$. As usual, the moduli integral in the string amplitude (\ref{eq:LLLLfull}) has divergences at $x\to 0,1,\infty$ that must be regularized. Here we employ the same regularization scheme as explained in \cite{Balthazar:2017mxh} by subtracting counter terms. Let us define
\ie{}
& R_s(x) = \sum_{0\leq n \leq (\omega_1+\omega_2)^2}a_n \int_{0}
^{\sqrt{(\omega_1+\omega_2)^2-n}}\frac{dP}{\pi} C^{s,s,s}(P,\omega_1,\omega_2)C^{s,s,s}(P,\omega_3,\omega_4)|x|^{-1+P^2-(\omega_1+\omega_2)^2+n},\\
& R_t(x) = \sum_{0\leq n \leq (\omega_1-\omega_3)^2}b_n \int_{0}^{\sqrt{(\omega_1-\omega_3)^2-n}}\frac{dP}{\pi} C^{s,s,s}(P,\omega_1,\omega_3)C^{s,s,s}(P,\omega_2,\omega_4)|1-x|^{-1+P^2-(\omega_1-\omega_3)^2+n},\\
& R_u(x) = x^{-2} \sum_{0\leq n \leq (\omega_1-\omega_4)^2}c_n \int_{0}^{\sqrt{(\omega_1-\omega_4)^2-n}}\frac{dP}{\pi} C^{s,s,s}(P,\omega_1,\omega_4)C^{s,s,s}(P,\omega_2,\omega_3)|1/x|^{-1+P^2-(\omega_1-\omega_4)^2+n},\\
\fe
where $a_0=b_0=c_0=1$, and the coefficients $a_n,b_n,c_n$ with $n\geq 1$ are chosen to cancel the part of $P$-integral in $s$, $t$, and $u$ channels that lead to divergences near $x=0, 1$, and $\infty$. These coefficients can be computed by expanding the prefactors and the conformal blocks in the moduli integrand as a power series in $x$. The regularized amplitude is written as
\ie\label{eq:LLLLfullreg}
&\cA_{\Psi^{s,s\,\,+}_{\omega_1}\Psi^{s,s\,\,+}_{\omega_2}\to\Psi^{s,s\,\,-}_{\omega_3}\Psi^{s,s\,\,-}_{\omega_4}} = \,\, i g_o^4C_{D^2}\delta(\omega_1+\omega_2-\omega_3-\omega_4)\\
&~~~~~\times \Bigg\{\sum_{j=1}^{3}\int_{I_j}dx\, |x|^{-2\omega_1\omega_2}|x-1|^{2\omega_2\omega_3} \langle \psi^{s,s}_{\omega_1}(0)\psi^{s,s}_{\omega_2}(x)\psi^{s,s}_{\omega_3}(1)\psi'^{s,s}_{\omega_4}(\infty) \rangle_{\text{Liouville}}
\\
&~~~~~~~~ - \int_{-\infty}^\infty dx \left[ R_s(x) + R_t(x) + R_u(x)\right] \Bigg\} +(3\leftrightarrow4).
\fe
It is understood here and below that the $x$-integral is performed after subtracting the counter terms at the level of integrands.

In the practical computation, we use crossing relations to map each moduli integration domain to $x\in(0,1/2)$. For instance, consider the diagram in Figure \ref{fig:LLLL}b, which corresponds to the integration domain $I_2=\{ 1< x < \infty \}$. We compute the Liouville correlator in the $\psi^{s,s}_{\omega_3}(1)\psi^{s,s}_{\omega_2}(x)$ OPE channel ($t$-channel) for $x\in(1,2)$, and in the $\psi^{s,s}_{\omega_2}(x)\psi^{s,s}_{\omega_4}(\infty)$ OPE channel ($u$-channel) for $x\in(2, \infty)$. We then change the integration variable $x\to 1-{1\over x}$ for the former and $x\to {1\over x}$ for the latter. In the end, the contribution to (\ref{eq:LLLLfull}) from Figure \ref{fig:LLLL}b, including the counter term, is given by
\ie\label{eq:LLLLdom}
&\cA'^{(I_2)}_{\Psi^{s,s\,\,+}_{\omega_1}\Psi^{s,s\,\,+}_{\omega_2}\to\Psi^{s,s\,\,-}_{\omega_3}\Psi^{s,s\,\,-}_{\omega_4}} \\
& = \int_0^{1/2} dx\,\bigg[  x^{2\omega_1\omega_3}(1-x)^{2\omega_1\omega_4}\int_0^\infty\frac{dP}{\pi} C^{s,s,s}(P,\omega_1,\omega_3)C^{s,s,s}(P,\omega_4,\omega_2)F(h_1,h_3,h_4,h_2;h\left| x\right.) \\
&~~~ +\,(1\leftrightarrow2)\vphantom{\frac{1}{1}} - {1\over x^2} R_t\left(1-{1\over x}\right) - {1\over x^2} R_u\left({1\over x}\right) \bigg] - \int_{1\over 2}^\infty {dx\over x^2} \left[ R_t\left(1-{1\over x}\right) + R_u\left({1\over x}\right) \right].
\fe
where $h_i=1+\omega_i^2$, $h=1+P^2$, and the prime denotes that we have excluded the prefactors in the first line of (\ref{eq:LLLLfull}). The other diagrams can be treated analogously.

\begin{figure}[h!]
\centering
 \subfloat[][]{\includegraphics[width=0.48\textwidth]{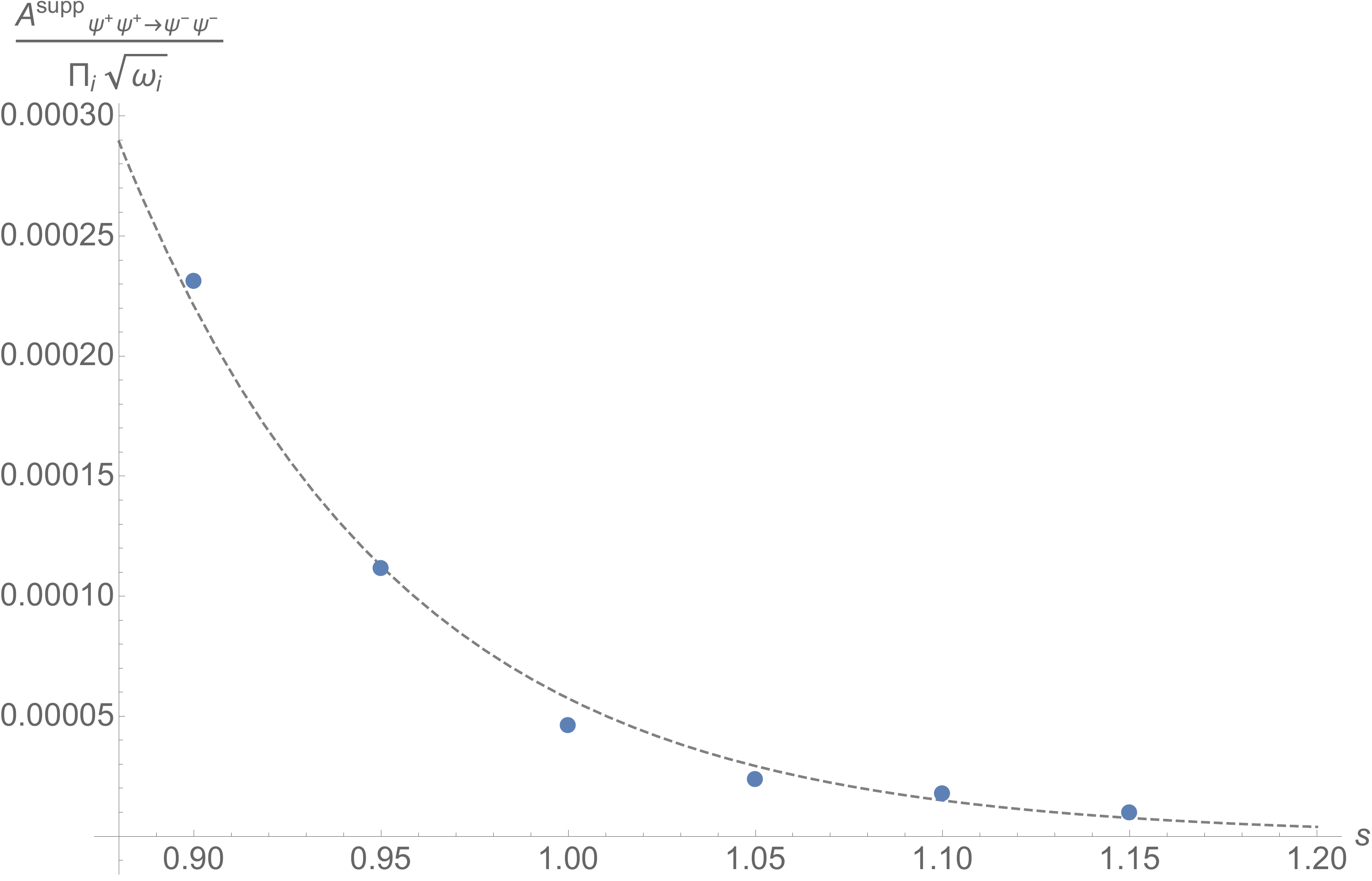}\label{fig:LLLL_supp}}~
 \subfloat[][]{\includegraphics[width=0.48\textwidth]{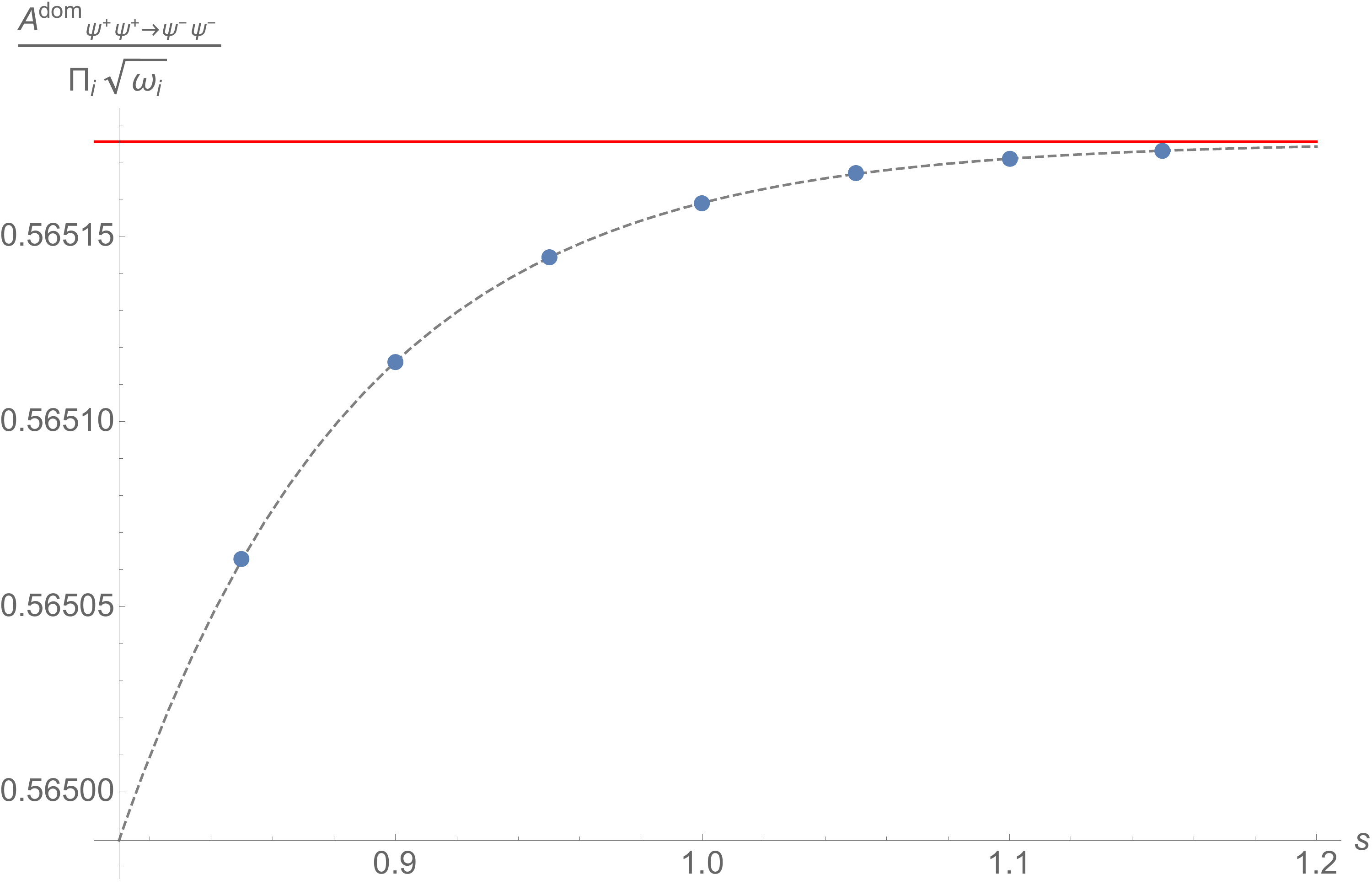}\label{fig:LLLL_dom}}
\caption{Contributions to the open string scattering amplitudes on FZZT brane from (a) the diagrams in Figures \ref{fig:LLLL}a,c and two other diagrams that are exponentially suppressed in the long string limit, and (b) the diagram in Figure \ref{fig:LLLL}b and another diagram that converge to $\cA_{L+L\to L+L}$ exponentially in the long string limit, evaluated numerically as a function of $s$, at fixed renormalized long string energies $\epsilon_1=0.2,\epsilon_2=0.4,\epsilon_3=0.35$. In comparison to (\ref{eq:LLLLfullreg}), we have included a factor of $1/\Pi_{i=1}^{4}\sqrt{\omega_i}$ due to the normalization of the long string asymptotic state. Exponential fits are shown as dashed gray curve, whereas the long string limit amplitude is marked in red.}
\label{fig:DomVsSupp}
\end{figure}

In fact, in the long string limit, the diagram corresponding to Figure \ref{fig:LLLL}b and the one related by exchanging ${P_3}\leftrightarrow{P_4}$ are the only ones that contribute to the scattering amplitude, whereas the four diagrams corresponding to Figures \ref{fig:LLLL}a,c and the ones related by ${P_3}\leftrightarrow{P_4}$ are exponentially suppressed. This is because the process depicted in Figure \ref{fig:LLLL}b involves a pair of long strings recombining in the region of spacetime where the effective string coupling is finite, whereas those of Figure \ref{fig:LLLL}a,c involves a pair of long strings joining their ends on the FZZT brane, where the effective string coupling is suppressed in the long string limit.

We have numerically evaluated all diagrams in Figure \ref{fig:LLLL} at fixed renormalized long string energies $\epsilon_i$, with increasing $s$. The results, as shown in Figure \ref{fig:DomVsSupp}, indeed demonstrate the expected exponential convergence in the long string limit. Note that the exponential suppression of Figure \ref{fig:LLLL}a,c results from a delicate cancelation including the regulators in the moduli integral. On the other hand, the surviving contributions, from Figure \ref{fig:LLLL}b and the one related by $P_3\leftrightarrow P_4$, are dominated by the Liouville momentum $P$-integral over a region away from where the regulator is needed, as was the case for the long $\to$ long $+$ closed amplitude. Further details are presented in Appendix \ref{sec:appnint}.

\begin{figure}[h!]
\centering
\includegraphics[width=10cm]{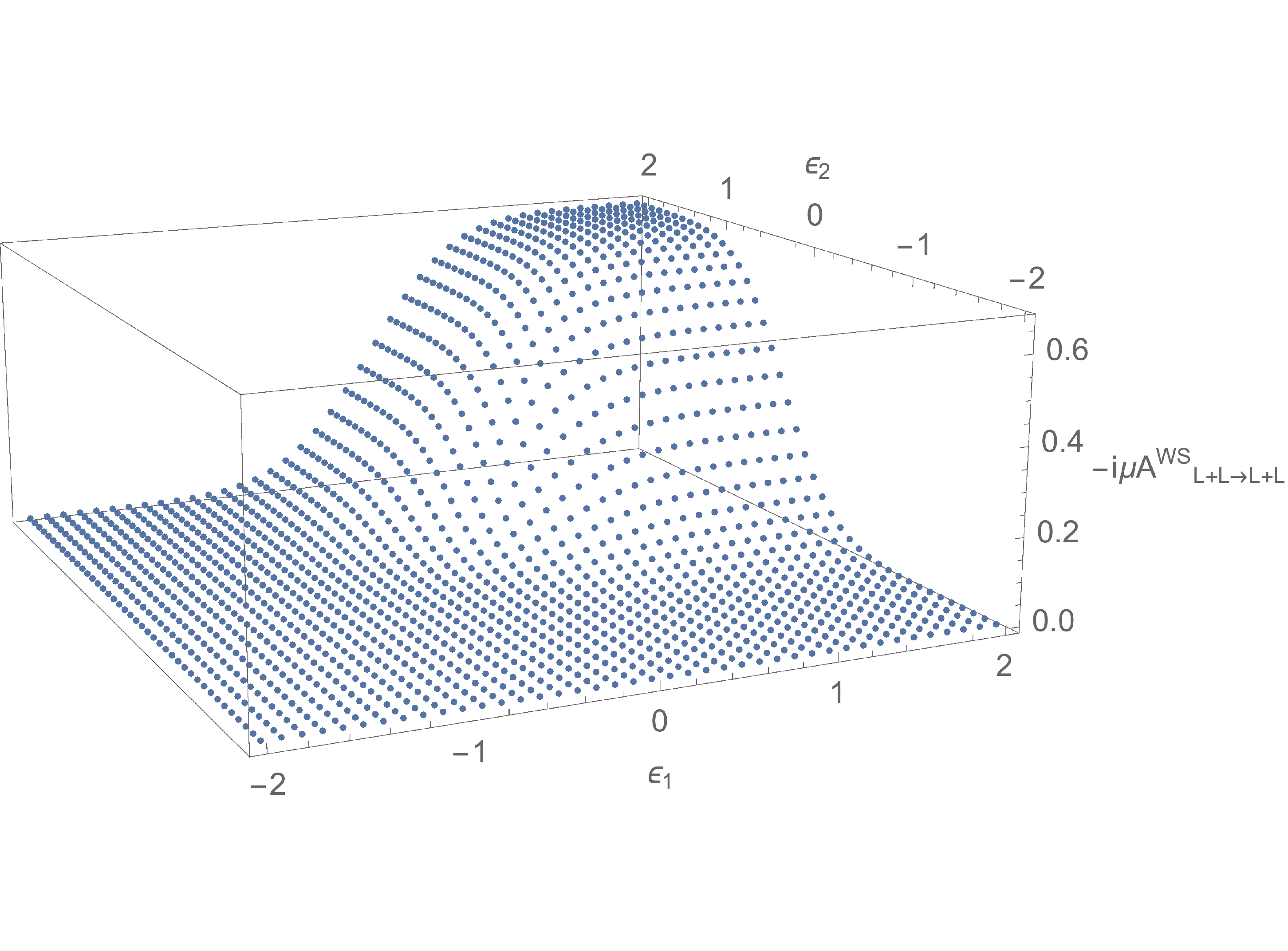}
\caption{Long string amplitude $\cA_{L+L\to L+L}$ as a function of incoming long string renormalized energies $\epsilon_1,\epsilon_2$, evaluated at a fixed outgoing long string renormalized energy $\epsilon_3=0.5$. 
}
\label{fig:LLLLvarye1e2}
\end{figure}

\begin{figure}[h!]
\centering
\subfloat[]{\includegraphics[width=0.45\textwidth]{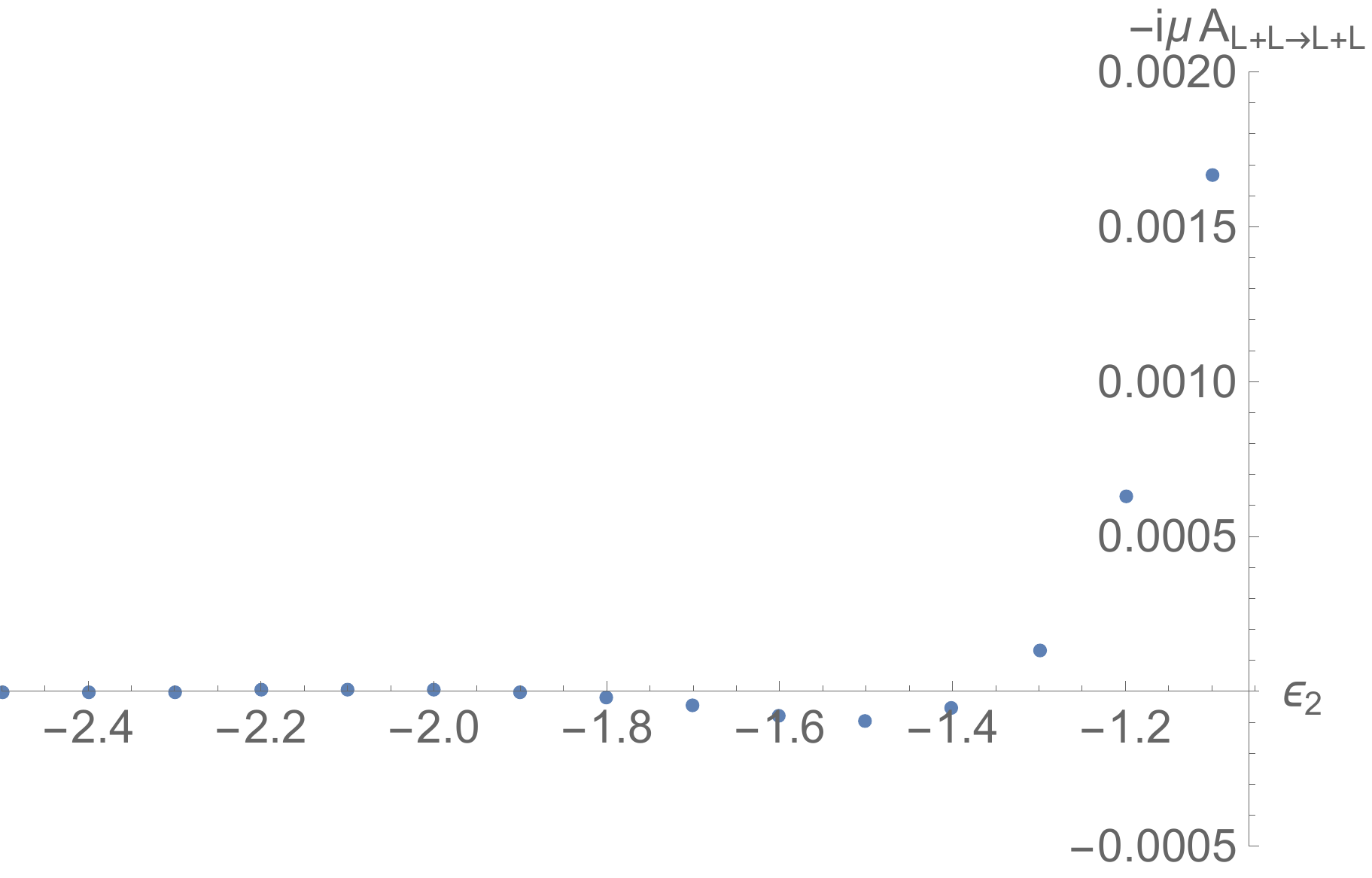}\label{fig:LLLLnege2dep}}~
\subfloat[]{\includegraphics[width=0.45\textwidth]{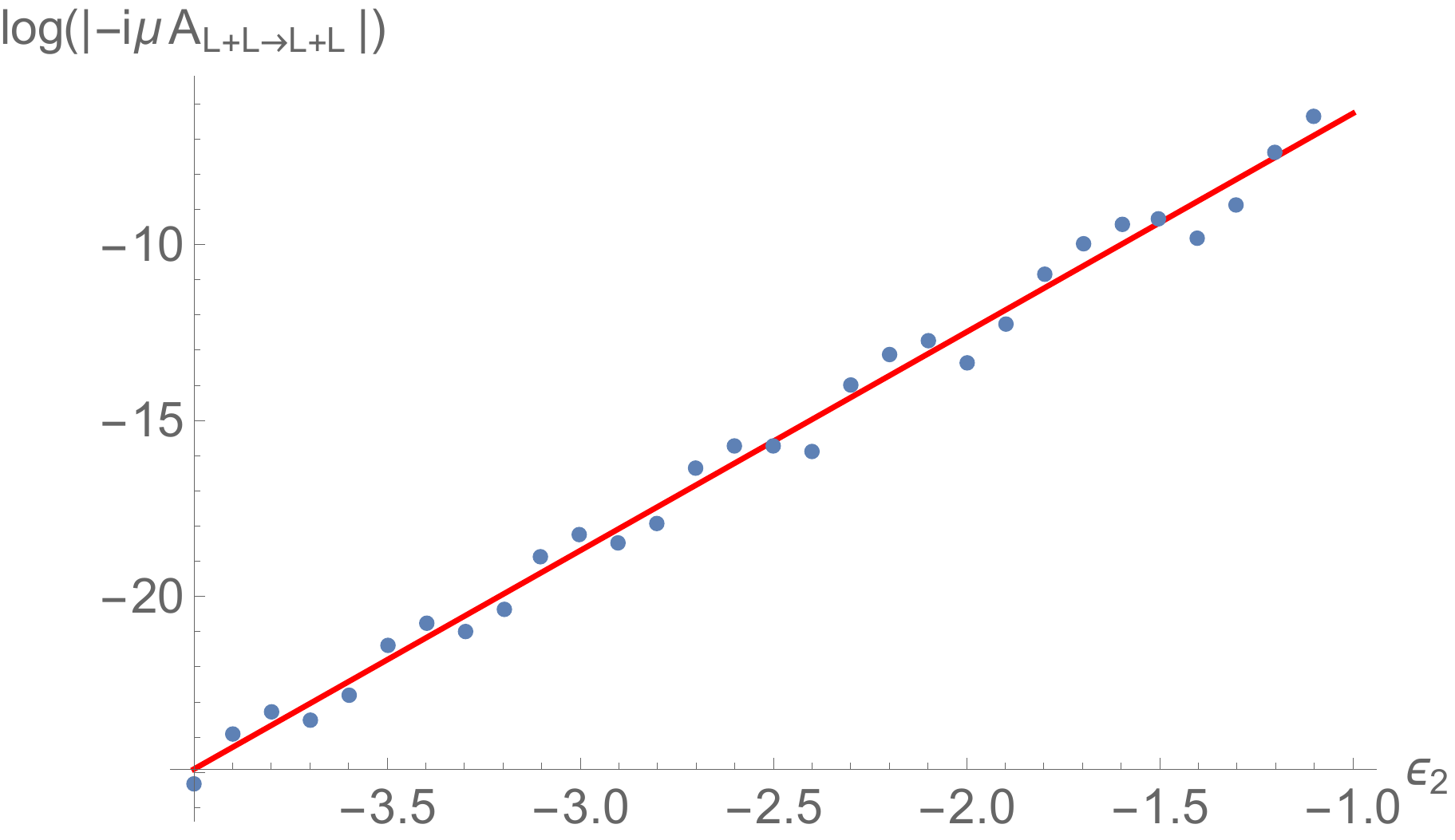}\label{fig:LLLLexpfitnege2}}
\caption{Long string amplitude $\cA_{L+L\to L+L}$ as a function of incoming long string renormalized energy $\epsilon_2$, evaluated at a fixed outgoing long string renormalized energies $\epsilon_1=3.0$ and $\epsilon_3=0.5$.  In (b) we plot the logarithm of the absolute value of the amplitude over a range of sufficiently large negative $\epsilon_2$. The red line represents a linear fit of slope $6.21$, which is in reasonable agreement with
 our expectation that the amplitude is modulated by a $e^{2\pi\mathrm{min}(\epsilon_i)}$ profile in this regime.
}
\label{fig:LLLL1d}
\end{figure}

In the end, we can evaluate $\cA_{L+L\to L+L}$ by an exponential fit of the numerical results of the open string amplitudes computed from Figure \ref{fig:LLLL}b and the diagram related by $P_3\leftrightarrow P_4$, without taking into account the counter terms while restricting the range of the $P$-integral. A sample of the results as a function of $\epsilon_1, \epsilon_2$, at a generic fixed value of $\epsilon_3$, is shown in Figure \ref{fig:LLLLvarye1e2}. We will see in section \ref{sec:LLLLMM} that this agrees with the corresponding amplitude computed in the dual matrix model to high numerical accuracy, which also allows us to determine the relation
\ie
\label{eq:go2}
g_o^2 = \frac{1}{2^{7/4}\sqrt{\pi} } g_s = \frac{1}{2^{11/4} \pi^\frac{3}{2} \,\mu}.
\fe

Some qualitative features of the results shown in Figures \ref{fig:LLLLvarye1e2} and \ref{fig:LLLL1d} can be understood semiclassically. For fixed $\epsilon_3$ and large $\epsilon_1, \epsilon_2$, the reconnecting of the pair of long strings occurs at the tip of the outgoing long string of least energy (namely $\epsilon_3$), where the effective string coupling is independent of $\epsilon_1, \epsilon_2$, which explains the plateau in Figure \ref{fig:LLLLvarye1e2}. For sufficiently large negative $\epsilon_i$, the amplitude is exponentially suppressed due to the suppression of the string coupling at the tip of the $i$-th long string, as shown in Figure \ref{fig:LLLL1d}.

\section{Non-singlet sectors of the matrix model}
\label{sec:MM}

The closed string sector of $c=1$ string theory, defined at the level of string perturbation theory, has long been conjectured to be dual to a suitable $N\to \infty$ limit of a $U(N)$-gauged Hermitian matrix quantum mechanics \cite{Klebanov:1991qa, Ginsparg:1993is, Jevicki:1993qn, McGreevy:2003kb, Martinec:2004td}. The closed string tree level and one-loop amplitudes computed from the worldsheet and matrix model descriptions are compared in detail in \cite{Balthazar:2017mxh} and convincing numerical agreement was found. 

The matrix model description of the long strings was proposed in \cite{Maldacena:2005hi} as certain states in non-singlet representations of the $U(N)$. Scattering amplitudes involving long strings in the matrix model have been previous explored in \cite{Karczmarek:2008sc, Bourgine:2007ua}. In this section we will first review the non-singlet sector of the matrix model, and the formulation of closed and long string scattering amplitudes at tree level. We will then perform the explicit computation of the tree amplitude of closed string emission by a long string, as well as the scattering of a pair of long strings, and compare the results numerically with the worldsheet computation.

\subsection{The Hamiltonian}

Let us begin with the quantum mechanical system consisting of an $N\times N$ Hermitian matrix variable $X$ and its conjugate canonical momentum matrix $P$ whose components are $P_{ij} = -i \partial/\partial X_{ji}$, with the Hamiltonian
\ie\label{hpxsys}
H = {\rm Tr} \left[{1\over 2} P^2+V(X)\right].
\fe
The potential $V(X)$ is given by $-{1\over 2} X^2$ in the domain of interest in the phase space. Any Hermitian matrix $X$ can be written as
\ie
X = \Omega^{-1} \Lambda \Omega
\fe
for some unitary matrix $\Omega$, where $\Lambda = {\rm diag} (\lambda_1, \cdots, \lambda_N)$ is diagonal. We can rewrite the wave function $\Psi(X)$ in the form
\ie
\Psi(X) \equiv \widehat\Psi(\Lambda, \Omega),
\fe
where $\widehat\Psi$ is now invariant under a $S_N\ltimes U(1)^N$ gauge redundancy. The $U(1)^N$ leaves $\Lambda$ invariant and acts on $\Omega$ by
\ie\label{zwrot}
\Omega \mapsto T^{-1} \Omega , ~~~ T = {\rm diag} (e^{i\A_1},\cdots, e^{i\A_N}).
\fe
The $S_N$ is generated by
\ie\label{snact}
\Lambda\mapsto W_{ij}^{-1} \Lambda W_{ij},~~~ \Omega\mapsto W_{ij}^{-1} \Omega ,
\fe
where $W_{ij}$, $i\neq j$, is the unitary matrix defined by 
\ie
(W_{ij})_{k\ell}=\begin{cases} 
      \delta_{ik} \delta_{j\ell} - \delta_{i \ell} \delta_{j k} & k=i~~ \mathrm{or}~~ k=j \\
       \delta_{k\ell} & \mathrm{otherwise} 
   \end{cases}
\fe
and acts on $\Lambda$ by permuting the pair of eigenvalues $\lambda_i$, $\lambda_j$. 

The Hamiltonian (\ref{hpxsys}) can be put in the form\footnote{It is convenient to use the identity
\ie
\sum_{k,l}(\Omega^{-1})_{ki}\Omega_{jl}{\partial\over \partial X_{kl}}=\begin{cases} 
      {\partial\over \partial\lambda_i} & i=j \\
       {R_{j i} \over \lambda_i - \lambda_j} & \mathrm{otherwise} 
   \end{cases}
\fe
where the derivatives act on functions of the Hermitian matrix $X$. This identity can be derived from the relation $\Omega dX \Omega^{-1} = d\Lambda + [\Lambda, d\Omega \,\Omega^{-1}]$.}
\ie\label{hamr}
H &=  \sum_{i=1}^N \left[ -{1\over 2}{\partial^2\over \partial \lambda_i^2} + V(\lambda_i) \right] + {1\over 2} \sum_{i\not= j} \left[ -{1\over \lambda_i - \lambda_j} {\partial\over \partial\lambda_i} + {R_{ij} R_{ji}\over (\lambda_i-\lambda_j)^2}\right]
\\
&= \Delta^{-1} H' \Delta ,
\fe
where $R_{\ell k} = \sum_m \Omega_{\ell m} {\partial\over \partial \Omega_{km}}$ is the $U(N)$ symmetry generator,
\ie
\Delta \equiv \prod_{i<j} (\lambda_i-\lambda_j),
\fe
and $H'$ is given by
\ie\label{hprime}
H' = \sum_{i=1}^N \left[ -{1\over 2} {\partial^2\over \partial \lambda_i^2} + V(\lambda_i) \right] + {1\over 2} \sum_{i\not= j} {R_{ij} R_{ji}\over (\lambda_i-\lambda_j)^2}.
\fe
If we redefine the wave function by
\ie\label{psirel}
\widehat\Psi(\Lambda, \Omega) \equiv \Delta^{-1} \Psi'(\Lambda, \Omega),
\fe
then $H'$ can be viewed as the Hamiltonian that acts on the wave function $\Psi'$.

The Hilbert space ${\cal H}$ of the system (\ref{hpxsys}) can be decomposed into sectors according to irreducible representations ${\cal R}$ of the $U(N)$ symmetry. We will take the point of view that the ${\cal R}$-sector of the matrix model is defined by the Hamiltonian (\ref{hamr}) acting on the Hilbert space
\ie\label{srihe}
{\cal H}_{\cal R} = \left[ L^2(\mathbb{R}^N)\otimes V^0_{\cal R}\right]^{S_N},
\fe
where $V^0_{\cal R}$ is the space of zero-weight states in the representation space of ${\cal R}$. The zero-weight condition is a consequence of the invariance of the wave function $\widehat\Psi(\Lambda, \Omega)$ under (\ref{zwrot}). The superscript $S_N$ in (\ref{srihe}) means to restrict to the subspace invariant under the $S_N$ action according to (\ref{snact}).

Equivalently, we can think of the ${\cal R}$-sector of the matrix model as being defined by the Hamiltonian (\ref{hprime}) acting on the space\footnote{Due to the relation (\ref{psirel}), $\widetilde{L^2}(\mathbb{R}^N)$ is the space of square-integrable functions defined with respect to the integration measure $d\mu = \prod_{k=1}^N d\lambda_k\, \Delta^{-2}$.}
\ie
{\cal H}_{\cal R}' = \left[ \widetilde{L^2}(\mathbb{R}^N)\otimes V^0_{\cal R}\right]^{S_N'},
\fe
where $S_N'$ acts on the wave function $\Psi'(\Lambda, \Omega)$ by (\ref{zwrot}) combined with an extra minus sign when a pair of eigenvalues are exchanged. 

The sector that contains $n$ long strings can be identified with the representation ${\cal R} = ({\rm adj})^{\otimes n}$. We will denote the corresponding Hilbert space ${\cal H}_{\cal R}'$ as simply ${\cal H}_n$. Here we will not view the long strings as identical particles, but rather as limits of open strings attached to $n$ different FZZT branes. ${\cal H}_n$ is spanned by states of the form
\ie\label{psiwv}
\psi_{i_1j_1, \cdots, i_n j_n}(\lambda_1, \cdots, \lambda_N) |i_1 j_1, \cdots, i_n j_n\rangle,
\fe
where the zero-weight condition means that $\{i_1, \cdots, i_n\}$ is a permutation of $\{j_1, \cdots, j_n\}$, and the statistics is such that $\psi$ is odd with respect to swapping $\lambda_i$ with $\lambda_j$ and at the same time swapping $i$ and $j$ among all of the indices of $\psi$. The $SU(N)$ generators act by
\ie
R_{ij} |k\ell\rangle = \delta_{jk} |i\ell\rangle - \delta_{i\ell} |kj\rangle.
\fe

\subsection{The singlet sector and collective field}
\label{singletsection}

We now briefly review the collective field description of the singlet sector of the matrix model, in the semiclassical limit of large eigenvalue density which corresponds to weak string coupling. The eigenvalues behave as $N$ free fermions subject to the Hamiltonian $H = {1\over 2} p^2 + V(\lambda)$, where $p$ is the canonical momentum conjugate to $\lambda$. The ground state is described by a fermi sea that fills in the region ${1\over 2} p^2 + V(\lambda) < -2\mu$ in the phase space. In particular, $V(\lambda) = -{1\over 2}\lambda^2$ in the domain of interest, and so $\lambda$ ranges from $\sqrt{2\mu}$ to infinity.

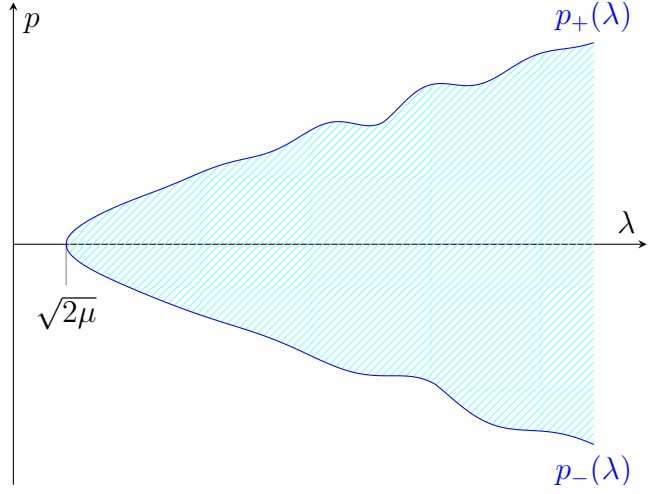
\begin{figure}[h]
\centering
{
\begin{tikzpicture}
\begin{axis}[axis lines=middle,
            xlabel=$\lambda$,
            ylabel=$p$,
            enlargelimits,
            ytick=\empty,
            xtick=\empty,
            height=8cm,
            width=10cm
           ]




\node[coordinate,pin=-90:{$\sqrt{2\mu}$}] at (axis cs:1,0){};

\addplot[name path=F,blue,domain={1:6},samples=750] {-{sqrt(x^2-1)*(1+0.1*exp(-sqrt((x-4.5)^2))*sin(4*deg(x)))}} node[pos=1, below]{$p_-(\lambda)$};

\addplot[name path=G,blue,domain={1:6},samples=750] {sqrt(x^2-1)*(1+0.1*exp(-sqrt((x-4)^2))*sin(6*deg(x)))}node[pos=1, above]{$p_+(\lambda)$};

\addplot[pattern=north east lines, pattern color=cyan!50]fill between[of=F and G, soft clip={domain=1:6}]
;

\end{axis}
\end{tikzpicture}

}
\caption{The low-energy dynamics of $N$ non-relativistic fermions can be described by a fluid in phase space. Low-energy excitations above the fermi surface are captured by deformations of the profile functions $p_\pm(\lambda)$.}
\label{fig:ppm}
\end{figure}

Low energy fluctuations can be characterized by those of the fermi surface $p=p_\pm(\lambda)$, see Figure \ref{fig:ppm}. 

The eigenvalue/fermion density is related by
\ie\label{eigdens}
\rho(\lambda) \equiv \sum_{i=1}^N \delta(\lambda-\lambda_i) = {1\over 2\pi} (p_+(\lambda) - p_-(\lambda)).
\fe
The ground state has $p_\pm = \pm \sqrt{\lambda^2-2\mu}$, and the fermion density $\rho_0(\lambda) = {1\over \pi}\sqrt{\lambda^2 - 2\mu}$. Deformations of the fermi surface can be characterized by density fluctuation
\ie
\rho(\lambda) = \rho_0(\lambda) + {1\over \sqrt\pi} \partial_\lambda \eta(\lambda).
\fe
The collective field $\eta(\lambda)$ has a conjugate momentum density
\ie
\Pi_\lambda = - {1\over 2\sqrt{\pi}} (p_+(\lambda) + p_-(\lambda))
\fe
that obeys the Poisson bracket $\{ \eta(\lambda), \Pi_\lambda(\lambda')\} = \delta(\lambda-\lambda')$. The free fermion Hamiltonian can be expressed in terms of the collective field as
\ie\label{ffhameta}
H = \int_{\sqrt{2\mu}}^\infty d\lambda \left[ {1\over 2}\sqrt{\lambda^2 - 2\mu} \left( \Pi_\lambda^2 + (\partial_\lambda\eta)^2 \right) 
+ {\sqrt{\pi}\over 2} (\Pi_\lambda)^2 \partial_\lambda\eta + {\sqrt{\pi}\over 6}(\partial_\lambda\eta)^3 \right].
\fe
It is convenient to pass to the coordinate $\tau$, related to $\lambda$ by $\lambda \equiv \sqrt{2\mu} \cosh\tau$. The canonical momentum density conjugate to $\eta(\tau)$ is $\Pi_\tau = \sqrt{2\mu} \sinh\tau \Pi_\lambda$. (\ref{ffhameta}) is written in $\tau$ coordinate as
\ie\label{ffhamtau}
H = \int_0^\infty d\tau \left[ {1\over 2} \left( \Pi_\tau^2 + (\partial_\tau\eta)^2 \right) 
+ {\sqrt{\pi}\over 12\mu (\sinh\tau)^2} \left( 3(\Pi_\tau)^2 \partial_\tau\eta + (\partial_\tau\eta)^3\right) \right].
\fe
Near the ``tip" of the fermi sea $\tau=0$, $\partial_\tau \eta$ is unconstrained and finite in the $\tau\to 0$ limit. This is compatible with a Dirichlet boundary condition $\dot\eta|_{\tau=0}=0$, or simply $\eta|_{\tau=0}=0$ as $\eta$ is only defined up to an additive constant. Note further that the interaction term in the Hamiltonian is singular at $\tau=0$. This can be treated \cite{Demeterfi:1991tz, Demeterfi:1991nw} by cutting off the $\lambda$-integral at $\lambda = \sqrt{2\mu} + \epsilon$ for some small positive parameter $\epsilon$, which amounts to cutting off $\tau$ at $\tau=\delta$ for $\delta = \sqrt{\epsilon} (\mu/2)^{-{1\over 4}}$. One then adds a local counter term that cancels against possible divergences that scale like inverse powers of $\epsilon$.

The closed string asymptotic states can be identified with that of $\eta(\tau)$. For instance, the tree level $1\to 2$ closed string scattering amplitude can be computed from (\ref{ffhamtau}) using the Born approximation\footnote{In our convention, the asymptotic mode expansion takes the form $\eta(\tau)=\int_0^\infty {dp\over \sqrt{\pi}}{1\over p}(b_p \sin p\tau +b_p^{\dagger}\sin p\tau)$, with $b_p, b_p^\dagger$ normalized according to $[b_p,b_{p'}^{\dagger}]=p\delta(p-p')$.},
\ie
S_{1\to 2}(\omega; \omega_1, \omega_2) &= \delta(\omega-\omega_1-\omega_2) {\cal A}_{1\to 2}(\omega_1, \omega_2)
\\
&= \delta(\omega-\omega_1-\omega_2){-i\omega \omega_1 \omega_2 \over \mu} \left[\int_\delta^\infty {d\tau \over (\sinh\tau)^2} + {\rm counter~term} \right].
\fe
The $\tau$-integral,
\ie
\int_\delta^\infty {d\tau \over (\sinh\tau)^2} = \coth\delta - 1,
\fe
simply gives $-1$ after taking into account the counter term that scales like $\delta^{-1}$, resulting in the amplitude ${\cal A}_{1\to 2} = {1\over \mu} i\omega \omega_1\omega_2$.

Alternatively, one may parameterize the fermi surface as a function of the momentum $p$, which ranges over the entire real axis, and express the excitations as those of a collective field that is a massless right-moving boson with cubic interaction, as discussed in Appendix \ref{pparasec}. In this description the interaction Lagrangian is non-singular and no UV regularization is required. Nonetheless, we find it more convenient to work with the collective field $\eta(\tau)$ in the non-singlet sector, when long strings are present.

\subsection{The long string state}

Now we turn to the adjoint sector and consider a state of the form
\ie\label{psiww}
\psi_{ij}(\lambda_1,\cdots, \lambda_N)|ij\rangle = \sum_{i=1}^N w(\lambda_i) \psi_0(\lambda_1, \cdots, \lambda_N)|ii\rangle \equiv |w\rangle,
\fe
where $\psi_0(\lambda_1,\cdots, \lambda_N)$ is the ground state wave function of the singlet sector of the matrix model. Note that, importantly, $w(\lambda_i)$ is viewed as a function $w(\lambda)$ evaluated at the eigenvalue $\lambda_i$, rather than a set of constant coefficients, so that $\psi_{ij}(\lambda_1,\cdots,\lambda_N)$ obeys the required statistics as described below (\ref{psiwv}). 

The Hamiltonian (\ref{hprime}) acts on (\ref{psiww}) as \cite{Marchesini:1979yq}
\ie\label{hprimact}
H' |w\rangle &= \sum_i \left[ -{1\over 2} w''(\lambda_i) - w'(\lambda_i) {\partial\over \partial \lambda_i} + E_0 w(\lambda_i) + \sum_{j\not=i} {w(\lambda_i) - w(\lambda_j) \over (\lambda_i - \lambda_j)^2} \right] \psi_0 |ii\rangle.
\fe
Here $E_0$ is the ground state energy of the singlet sector, and will be omitted from now. In the weak string coupling limit, corresponding to the limit of large eigenvalue density, the last term in the bracket on the RHS of (\ref{hprimact}) dominates. In terms of the eigenvalue density (\ref{eigdens}), we may also write this term as
\ie
\int d\lambda\,\rho(\lambda) {w(\lambda_i) - w(\lambda)\over (\lambda_i-\lambda)^2} \psi_0 |ii\rangle.
\fe
According to the proposal of \cite{Maldacena:2005hi}, the single long string asymptotic state of energy $E$ is given by (\ref{psiww}) with $w(\lambda)=w_E(\lambda)$ obeying the eigen-energy equation
\ie\label{lstreqa}
& \int_{\sqrt{2\mu}}^\infty d\lambda'\,\rho_0(\lambda') {w_E(\lambda) - w_E(\lambda')\over (\lambda-\lambda')^2} = E w(\lambda)
\fe
and the normalization condition 
\ie\label{lstreqb}
& \int_{\sqrt{2\mu}}^\infty d\lambda\, \rho_0(\lambda) w_E(\lambda) w_{E'}^*(\lambda) = \delta(E-E'),
\fe
where $\rho_0(\lambda)={1\over \pi} \sqrt{\lambda^2-2\mu}$ is the ground state eigenvalue density in the singlet sector. The long string energy $E$ is subject to an infrared divergence: if we cut off $\tau$ at a large distance $L$, or $\lambda$ at $\sqrt{2\mu} \cosh L$, then $E$ diverges like ${L\over \pi}$. The renormalized long string energy, defined by (\ref{eq:lslimit}) in the worldsheet description, will be identified as
\ie\label{mmrenormenerg}
\epsilon = E - {L-1\over \pi}
\fe
in the matrix model description. We have determined the finite shift in (\ref{mmrenormenerg}) a posteriori by the numerical comparison of long string amplitudes on the two sides of the duality in sections \ref{sec:LLCMM} and \ref{sec:LLLLMM}.
It is convenient to pass from $\lambda$ to $\tau$ coordinate, and write $h_\epsilon(\tau)\equiv \sqrt{\pi} \rho_0(\lambda) w_E(\lambda)$. (\ref{lstreqa}) and (\ref{lstreqb}) can be rewritten as
\ie{}
& {1\over 4\pi} \int_0^\infty d\tau' \left[ {1\over \left(\sinh {\tau+\tau'\over 2}\right)^2} - {1\over \left(\sinh {\tau-\tau'\over 2}\right)^2} \right] h_\epsilon(\tau') - {1\over \pi} {\tau\over \tanh\tau} h_\epsilon(\tau) = \epsilon \, h_\epsilon(\tau),
\\
& \int_0^\infty d\tau \, h_\epsilon(\tau) h_{\epsilon'}(\tau) = \delta(\epsilon - \epsilon').
\fe
Note that the integrand in the first line is a priori singular at $\tau'=\tau$, and is defined by principal value prescription. Furthermore, $h_\epsilon(\tau)$ is subject to the boundary condition $h_\epsilon(\tau=0)=0$. The exact solutions are found in \cite{Fidkowski:2005ck}, 
\ie
\label{Fidkwavef}
h_\epsilon(\tau) = \int_{-\infty}^\infty {dk\over \sqrt{\pi}} {\sin(k\tau) \sinh(\pi k)\over \sqrt{\sinh^2(\pi k) + e^{2\pi \epsilon}}}\, \sin\left[\pi \int_{k_0}^k {dk' \,\sinh(\pi k')\over \sqrt{\sinh^2(\pi k') + e^{2\pi \epsilon}}} \left( \epsilon - { k'\over \tanh(\pi k')} \right) \right],
\fe
where $k_0 = {i\over \pi} \arcsin e^{\pi\epsilon}$.

\subsection{The long $\to$ long $+$ closed amplitude in the adjoint sector}
\label{sec:LLCMM}

Let us begin with the long string asymptotic state $|w\rangle$, which is an eigenstate of the ``free" part of the Hamiltonian $H'$. The interaction part of $H'$ acts on $|w\rangle$ as
\ie
H_{int}' |w\rangle = \sum_i \left[ -{1\over 2} w''(\lambda_i) - w'(\lambda_i) {\partial\over \partial \lambda_i} + {1\over \sqrt{\pi}} \int d\lambda\,\partial_\lambda\eta (\lambda) {w(\lambda_i) - w(\lambda)\over (\lambda_i-\lambda)^2}  \right] \psi_0 |ii\rangle.
\fe
Let $b_\omega, b_\omega^\dagger$ be the out-state annihilation and creation operators of a closed string, i.e. of a mode of the collective field $\eta(\tau)$ of energy $\omega$. An out-state of a long string and a closed string can be represented as $b_\omega^\dagger |w_E\rangle$.

We would like to compute the amplitude of a long string of energy $E_1$ decaying into a long string of energy $E_2$ and a closed string of energy $\omega_3$, with $E_1=E_2+\omega_3$. At tree level, this is given by the Born approximation
\ie{}
&{\cal A}_{L\to L+C}^{\rm tree} = -2\pi i \langle w_{E_2} | b_{\omega_3} H_{int}' |w_{E_1}\rangle 
\\
&=  2\pi^{3\over 2} \int d\lambda \, \rho_0(\lambda)  \langle\psi_0|b_{\omega_3} \Pi_\lambda(\lambda)|\psi_0\rangle w_{E_2}^*(\lambda) \partial_\lambda w_{E_1}(\lambda) 
\\
&~~~~ -  2 i \sqrt{\pi} \int d\lambda d\lambda' \, \rho_0(\lambda) \langle\psi_0|b_{\omega_3} \partial_{\lambda'}\eta(\lambda') |\psi_0\rangle   {w_{E_2}^*(\lambda) ( w_{E_1}(\lambda) - w_{E_1}(\lambda') ) \over (\lambda-\lambda')^2}.
\fe
Rewriting in $\tau$ coordinate, we have
\ie
{\cal A}_{L\to L+C}^{\rm tree} &= {i\over\mu} \left[\pi\omega_3 \int_0^\infty d\tau \,\sin(\omega_3 \tau) {h_{\epsilon_2}^*(\tau)\over \sinh\tau} \partial_\tau\left(  {h_{\epsilon_1}(\tau)\over \sinh\tau} \right)\right.
\\
&~~~~~~~\left. - \omega_3 \int_0^\infty d\tau d\tau' { \cos(\omega_3 \tau') \over (\cosh \tau - \cosh\tau')^2} h_{\epsilon_2}^*(\tau) \left( h_{\epsilon_1}(\tau) - \frac{\sinh\tau}{\sinh\tau'} \, h_{\epsilon_1}(\tau') \right)\right].
\label{eq:LLCampMM}
\fe

This amplitude was also found in \cite{Karczmarek:2008sc,Donos:2005vm} where a collective field for the adjoint sector is introduced.

\begin{figure}[h!] 
\centering
\subfloat[] {\includegraphics[width=0.7\textwidth]{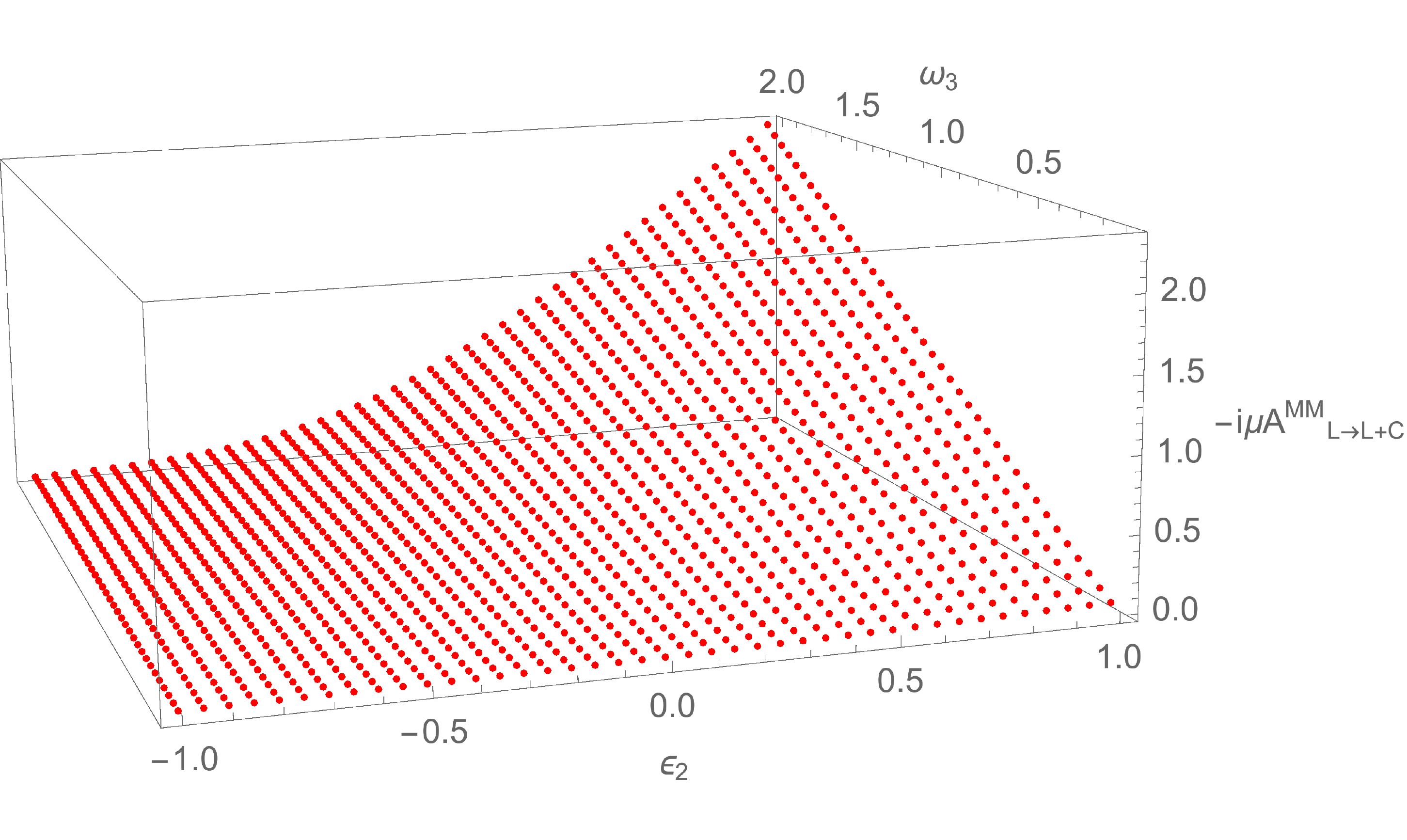}}\\
 \subfloat[][$\omega_3=1$.]{\includegraphics[width=0.48\textwidth]{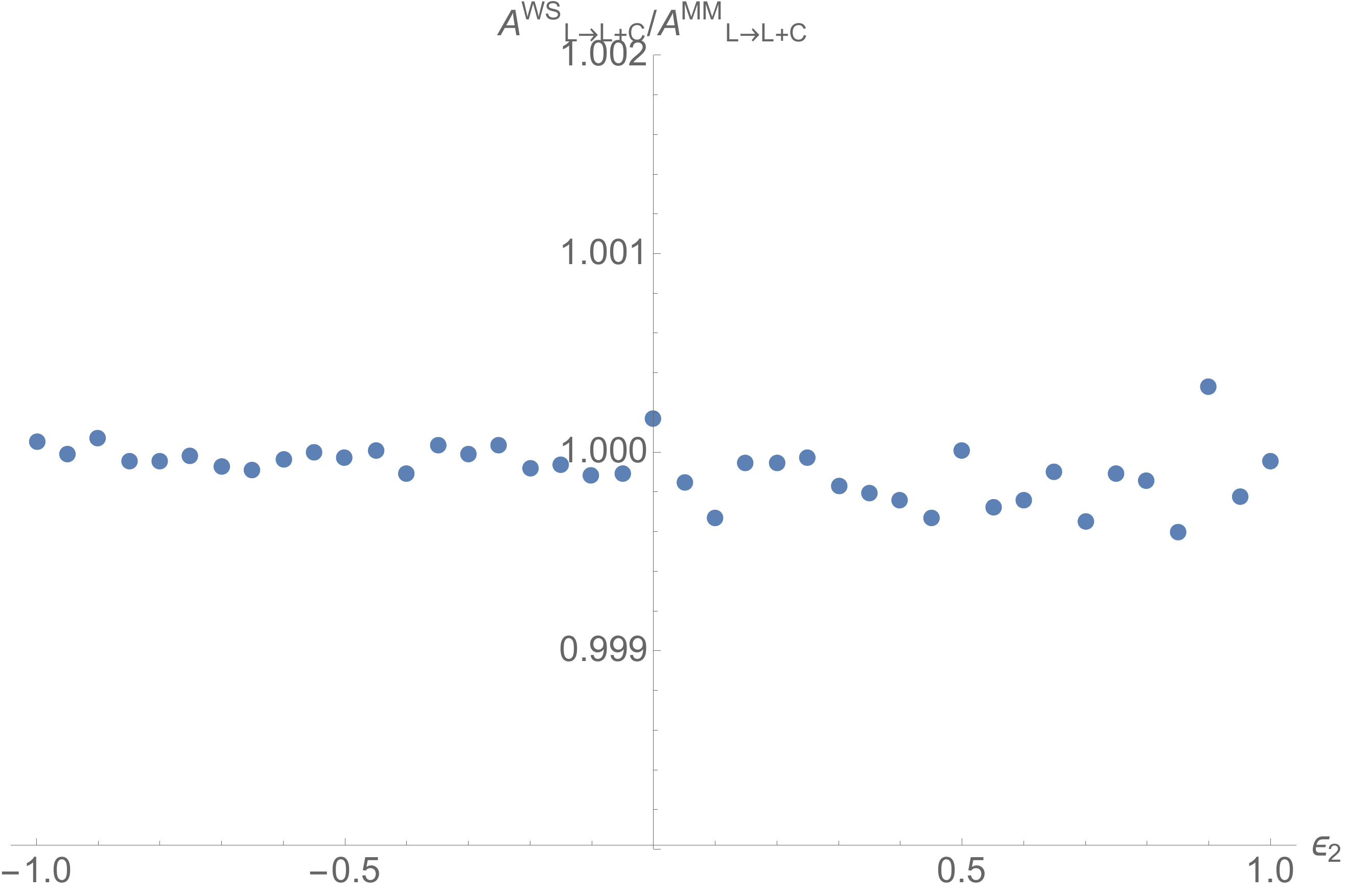}}~
 \subfloat[][$\epsilon_2=1/2$.]{\includegraphics[width=0.48\textwidth]{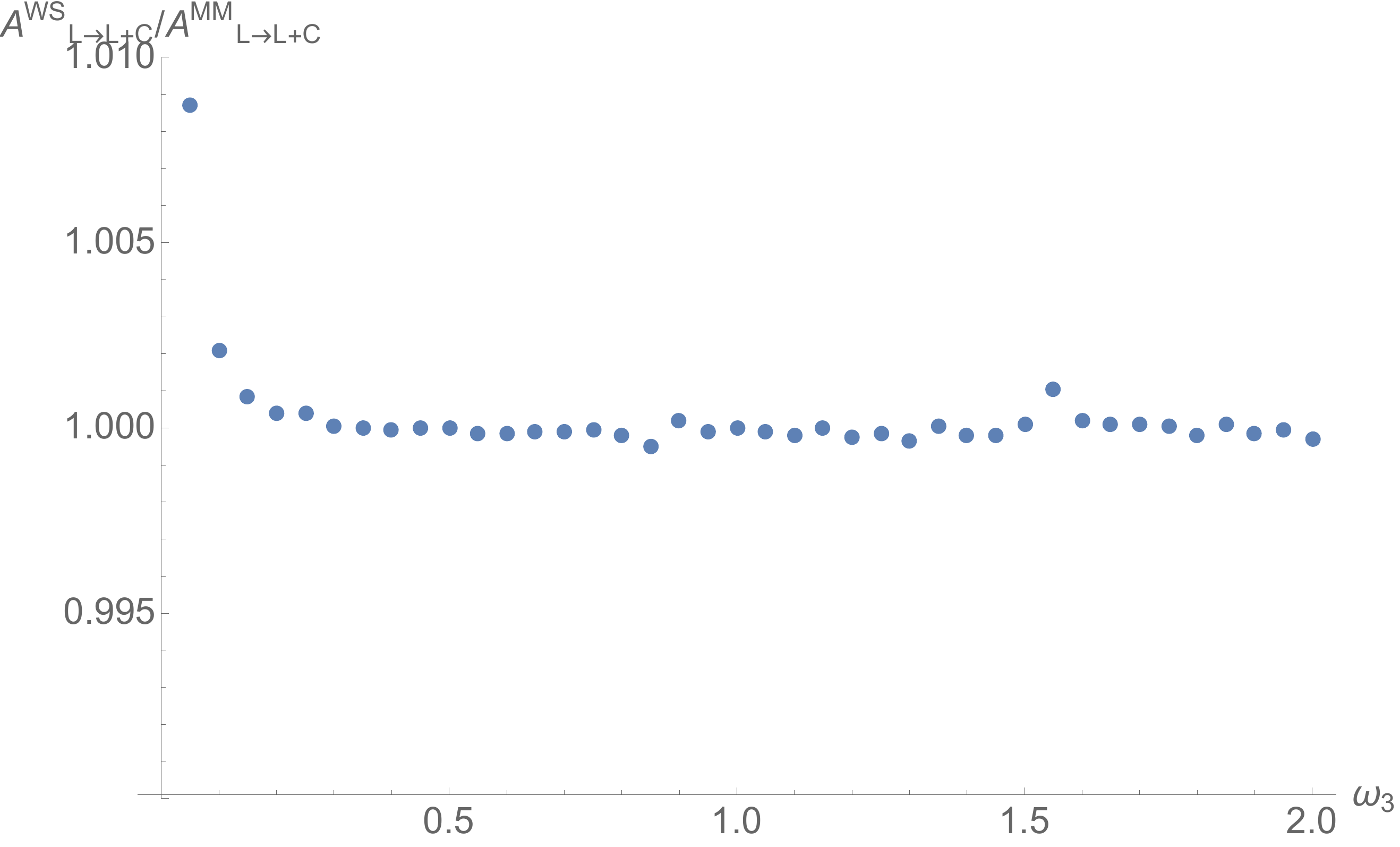}}
\caption{(a): Numerical results for the matrix model scattering of the tree level long $\to$ long $+$ closed amplitude (\ref{eq:LLCampMM}) as a function of the outgoing long string renormalized energy $\epsilon_1$ and outgoing closed string energy $\omega_3$. (b) and (c): Ratio of numerical results computed from the worldsheet to those computed from the matrix model for tree-level long $\to$ long $+$ closed string amplitude.}
\label{fig:MMampLLC}
\end{figure}

We evaluate the integrals in (\ref{eq:LLCampMM}) numerically, and the resulting amplitudes are shown in Figure \ref{fig:MMampLLC}. The result, up to an overall normalization constant, is in striking agreement with the worldsheet computation presented in Figure \ref{fig:LLCampdep}, up to $< 0.8\% $ error. By demanding that the overall normalizations agree as well, we have fixed $C_{D^2}$ in terms of $g_o$ with the result given in (\ref{eq:CD2}). Assuming (\ref{eq:CD2}), the ratio of the worldsheet and matrix model amplitudes as a function either of the long string energy $\epsilon_2$ or of the closed string energy $\omega_3$ is shown in Figure \ref{fig:MMampLLC}.

\subsection{The long $+$ long $\to$ long $+$ long amplitude in the bi-adjoint sector}
\label{sec:LLLLMM}

The asymptotic state of a pair of long strings of energies $E_1$ and $E_2$ come in two types,
\ie
\sum_{i,j=1}^N w_{E_1} (\lambda_i) w_{E_2}(\lambda_j) \psi_0(\lambda_1, \cdots, \lambda_N) |ii, jj\rangle \equiv |w_{E_1}, w_{E_2}\rangle_{11,22},
\fe
and
\ie
\sum_{i,j=1}^N w_{E_1} (\lambda_i) w_{E_2}(\lambda_j) \psi_0(\lambda_1, \cdots, \lambda_N) |ij, ji\rangle \equiv |w_{E_1}, w_{E_2}\rangle_{12,21},
\fe
The former has the interpretation of a pair of long strings each of which has both ends on the same FZZT brane (labeled by either 1 or 2), whereas the latter describes a pair of long strings connecting two FZZT branes, with opposite orientations. 

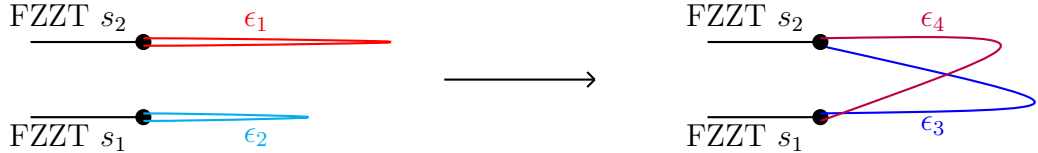
\begin{figure}[h!]
\centering
\begin{tikzpicture}

\draw[fill] (1,0) circle [radius=0.1];
\draw [thick]  (-0.5,0) -- (1,0);
\draw (0,0.05) node[below] {FZZT $s_1$};
\draw [-,thick, cyan] (1,0.05) to [out=0,in=0] (3,0)
to [out=0,in=0] (1,-0.05) ;
\draw (2.5,1) node[above,red] {$\epsilon_1$};

\draw[fill] (1,1) circle [radius=0.1];
\draw [thick] (-0.5,1) -- (1,1);
\draw (0,1) node[above] {FZZT $s_2$}; 
\draw [-,thick, red] (1,1.05) to [out=0,in=0] (4,1)
to [out=0,in=0] (1,0.95) ;
\draw (2.5,0) node[below,cyan] {$\epsilon_2$};

\draw [thick] (5,0.5) -- (7,0.5);
\draw [thick] (7,0.5) -- (6.9,0.6);
\draw [thick] (7,0.5) -- (6.9,0.4);

\draw[fill] (10,0) circle [radius=0.1];
\draw [thick]  (8.5,0) -- (10,0);
\draw (9,0.05) node[below] {FZZT $s_1$};
\draw [-,thick, blue]  (10,0.05) .. controls (13.8,0.1) .. (10,0.95) ;
\draw (11.5,1) node[above,purple] {$\epsilon_4$};

\draw[fill] (10,1) circle [radius=0.1];
\draw [thick] (8.5,1) -- (10,1);
\draw (9,1) node[above] {FZZT $s_2$}; 
\draw [-,thick, purple]  (10,1.05) .. controls (13.2,1.1) .. (10,-0.05) ;
\draw (11.5,0.15) node[below,blue] {$\epsilon_3$};

\end{tikzpicture}
\caption{The $11,22\to 12,21$ long string scattering by reconnecting.}
\end{figure}

We are interested in the tree level amplitude of a pair of long strings that interact by reconnecting in the bulk, thereby turning a state of type $11, 22$ to a state of type $12, 21$. It can be computed in the Born approximation as
\ie{}
&{\cal A}^{\rm tree}_{L+L\to L+L} = - 2\pi i {}_{12,21}\langle w_{E_3}, w_{E_4}|  \left[ {1\over 2} \sum_{i\not=j} {R_{ij}R_{ji}\over (\lambda_i - \lambda_j)^2} - (E_1 + E_2) \right] |w_{E_1}, w_{E_2}\rangle_{11,22}
\\
&= 2\pi i \int d\lambda d\lambda' \rho_0(\lambda) \rho_0(\lambda') {w_{E_3}^*(\lambda) w_{E_4}^*(\lambda') (w_{E_1}(\lambda) -w_{E_1}(\lambda') ) (w_{E_2}(\lambda) -w_{E_2}(\lambda') )\over (\lambda - \lambda')^2}
\\
& = {\pi i \over \mu} \int_0^\infty {d\tau d\tau'\,\sinh\tau \sinh\tau' \over (\cosh\tau - \cosh\tau')^2} h_{\epsilon_3}^*(\tau) h_{\epsilon_4}^*(\tau')
\left( {h_{\epsilon_1}(\tau)\over \sinh\tau} - {h_{\epsilon_1}(\tau')\over \sinh\tau'}\right) \left( {h_{\epsilon_2}(\tau)\over \sinh\tau} - {h_{\epsilon_2}(\tau')\over \sinh\tau'}\right).
\label{eq:LLLLampMM}
\fe
Note that at this order, the $11,22\to 11,22$ and $12,21\to 12,21$ amplitudes vanish identically.




\begin{figure}[h!]
\centering
\subfloat[]{\includegraphics[width=0.48\textwidth]{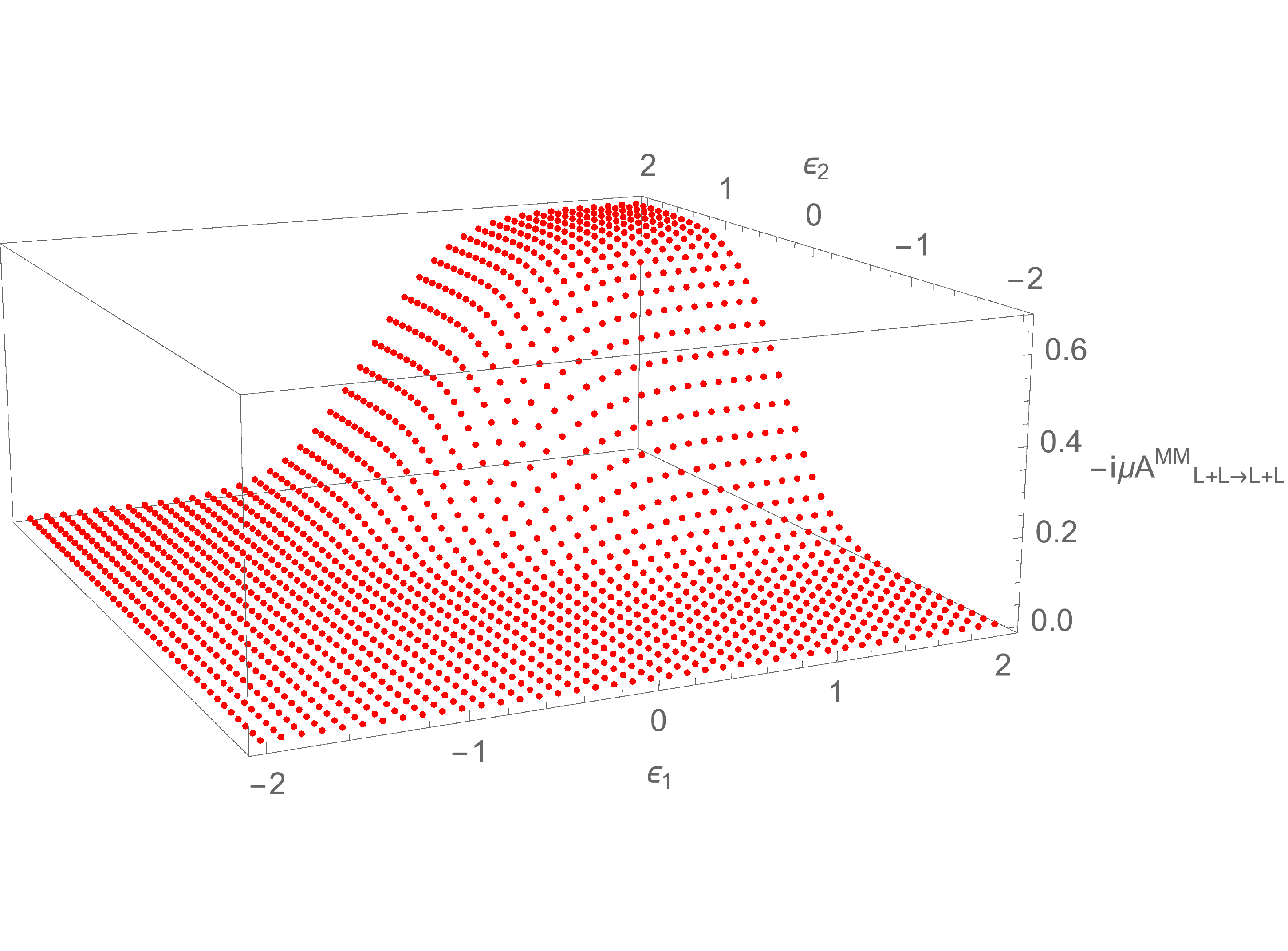}}~
\subfloat[][$\epsilon_2=0.5$.]{\includegraphics[width=0.48\textwidth]{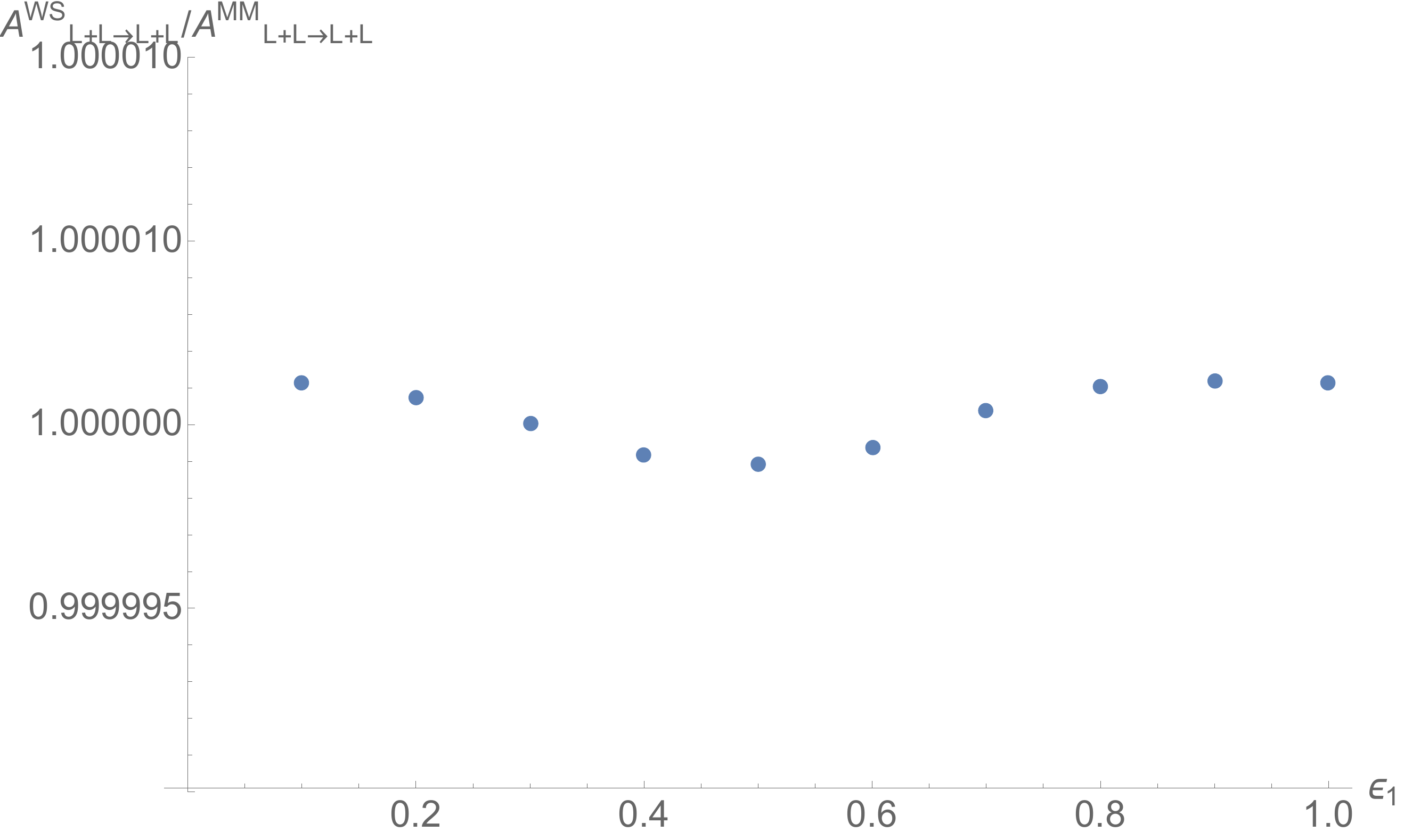}}
\caption{(a): Numerical results for the matrix model scattering of the tree level long $+$ long $\to$ long $+$ long amplitude (\ref{eq:LLLLampMM}) as a function of incoming long string renormalized energies $\epsilon_1,\epsilon_2$ at fixed outgoing long string renormalized energy $\epsilon_3=0.5$. (b): Ratio of numerical results computed from the worldsheet to those computed from the matrix model for tree-level long $+$ long $\to$ long $+$ long string amplitude.}
\label{fig:MMampLLLL}
\end{figure}

We evaluate the integrals in (\ref{eq:LLLLampMM}) numerically, and show the results in Figure \ref{fig:MMampLLLL}. Up to an overall normalization constant, the results are in excellent agreement with the worldsheet computation presented in Figure \ref{fig:LLLLvarye1e2}, up to $< 0.03\% $ error. By demanding the agreement of the overall normalization as well, we have fixed the relation between $g_o$ and $1/\mu$, or between $g_o$ and $g_s$, as given in (\ref{eq:go2}). A sample of the ratio of the worldsheet and matrix model results, as a function of one of the long string energies, is shown in Figure \ref{fig:MMampLLLL}.

\section{Discussion}
\label{sec:conc}

The main results of this paper are the computations of tree level amplitudes in $c=1$ string theory of closed string emission by a long string, and the scattering of a pair of long strings, both from the worldsheet description of long strings as limits of high energy open strings on receding FZZT branes, and from the matrix model description as certain states in non-singlet sectors of the matrix quantum mechanics. The worldsheet computation requires numerically integrating Virasoro conformal blocks and Liouville structure constants on the disc. The matrix model computation is expressed in terms of Fidkowski's long string solution \cite{Fidkowski:2005ck} and is evaluated numerically as well. Results of two sides of are found to be in striking agreement, giving strong support of the duality conjectured in \cite{Maldacena:2005hi}.

As both the long string and closed string behave like massless relativistic particles in the asymptotic region, one may worry about whether the S-matrix is well defined. Indeed certain closed string amplitudes, such as the tree level $2\to 2$ amplitude, are subject to ambiguities in the definition of asymptotic states, which is also reflected in non-analyticity (discontinuity) of the amplitudes across real energies. Such non-analyticity of the amplitude is physical and is related to intermediate on-shell particles \cite{Balthazar:2017mxh}. A similar discontinuity is expected of the tree level $2\to 2$ amplitude of open strings on FZZT branes, but should be absent in the long string limit, since the $2\to 2$ amplitude of long string does not factorize through an open string channel. Indeed, numerical evaluation of ${\cal A}^{\rm tree}_{LL\to LL}$ at complex energies suggest that the amplitude is in fact analytic across real energies. It should be possible to prove this analytically.

One of the outstanding open questions in the subject of $c=1$ string duality is the role of FZZT branes in the matrix model. It was argued in \cite{Gaiotto:2005gd} that the answer should be provided by coupling the matrices to new degrees of freedom that transform in the fundamental representation of the $U(N)$ (see also \cite{Betzios:2017yms}). While such a proposal may ultimately be correct, it remains to be clarified how the parameters of FZZT branes and the open string states are represented in the matrix model. We hope to report on these questions in the future.



\section*{Acknowledgments}

We are grateful to Minjae Cho, Scott Collier, Davide Gaiotto, Joanna Karczmarek, Igor Klebanov, Shota Komatsu, Juan Maldacena, Gregory Moore and Washington Taylor for discussions, and to Juan Maldacena for comments on a preliminary draft. We thank the organizers of Strings 2018, Bootstrap 2018, and the high energy theory group at Caltech for their hospitality during the course of this work. This work is supported in part by a Simons Investigator Award from the Simons Foundation, by the Simons Collaboration Grant on the Non-Perturbative Bootstrap, and by DOE grant DE-FG02-91ER40654. VR is supported by the National Science Foundation Graduate Research Fellowship under Grant No. DGE1144152. BB is supported by the Bolsa de Doutoramento FCT fellowship. A part of the numerical computation in this work is performed on the Odyssey cluster supported by the FAS Division of Science, Research Computing Group at Harvard University.

\appendix

\section{Verification of crossing relation in boundary Liouville correlators}
\label{sec:appcross}

In this section we check numerically that the boundary Liouville structure constants introduced in section \ref{fzztbraneintro} obey crossing relations at the level of disc bulk 2-point functions and disc boundary 4-point functions.

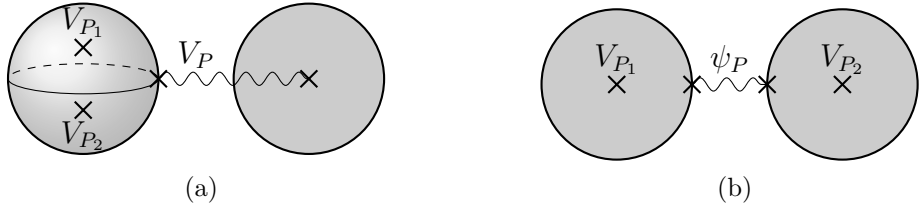
\begin{figure}[h!]
\centering
\subfloat[]{~~~~~
\begin{tikzpicture}
\shade[ball color = gray!40, opacity = 0.4] (0,0) circle (1);
  \draw[thick] (0,0) circle (1);
  \draw (-1,0) arc (180:360:1 and 0.2);
  \draw[dashed] (1,0) arc (0:180:1 and 0.2);
  \draw (0,5/12) node[cross=4pt, thick] {};
  \draw (0,5/12) node[above] {$V_{P_1}$};
  \draw (0,-5/12) node[cross=4pt, thick] {};
   \draw (0,-5/12) node[below] {$V_{P_2}$};
  \draw (1,0) node[cross=4pt, thick] {};

\filldraw[color=black, fill=gray!40, thick] (3,0) circle (1);
\draw (3,0) node[cross=4pt, thick] {};

\draw (1.5,0) node[above] {$V_{P}$};
\path [draw=black,snake it]
    (1,0) -- (3,0);
\end{tikzpicture}
\label{fig:disk2ptCA}~~~~~
}~~\subfloat[]{~~~~~
\begin{tikzpicture}
\filldraw[color=black, fill=gray!40, thick] (0,0) circle (1);
\draw (0,0) node[cross=4pt, thick] {};
\draw (0,0) node[above] {$V_{P_1}$};
\draw (1,0) node[cross=4pt, thick] {};

\draw (1.5,0) node[above] {$\psi_{P}$};
\path [draw=black,snake it]
    (1,0) -- (2,0);

\filldraw[color=black, fill=gray!40, thick] (3,0) circle (1);
\draw (3,0) node[cross=4pt, thick] {};
\draw (3,0) node[above] {$V_{P_2}$};
\draw (2,0) node[cross=4pt, thick] {};
\end{tikzpicture}
\label{fig:disk2ptRR}
~~~~~
}
\caption{(a) The OPE channel of disc bulk two-point function. (b) The boundary channel of disc bulk two-point function.}
\end{figure}

First consider the correlation function of a pair of bulk operators $V_{P_1}$ and $V_{P_2}$ in $c=25$ Liouville theory on the disc, subject to FZZT boundary condition labeled by the parameter $s$. This correlator admits two different Virasoro conformal block decompositions, which we refer to as the OPE channel and the boundary channel respectively, as illustrated in Figure \ref{fig:disk2ptCA} and \ref{fig:disk2ptRR}. In the OPE channel (Figure \ref{fig:disk2ptCA}), the disc correlator is expressed in terms of the bulk (DOZZ) structure constant and the disc one-point function,
\ie\label{discopechann}
&\left\langle V_{P_1}(z_1,\bar{z}_1) V_{P_2}(z_2,\bar{z}_2) \right\rangle_{D^2,s}=\left|z_1-\bar{z}_2\right|^{-4(1+P_1^2)}\left|z_2-\bar{z}_2\right|^{2(P_1^2-P_2^2)}\\
&\int_0^\infty \frac{dP}{\pi}A^s(P)\mathcal{C}(P_1,P_2,P) F\left(1+P_1^2,1+P_2^2,1+P_1^2,1+P_2^2;1+P^2\left|\frac{|z_1-z_2|^2}{|z_1-\bar{z}_2|^2}\right.\right),
\fe
where the relevant disc Virasoro conformal block can be identified with the sphere 4-point holomorphic Virasoro conformal block evaluated at a real cross ratio.

In the boundary channel (Figure \ref{fig:disk2ptRR}), the correlator is written in terms of disc bulk-boundary structure constants,
\ie\label{discbdrychann}
&\left\langle V_{P_1}(z_1,\bar{z}_1) V_{P_2}(z_2,\bar{z}_2) \right\rangle_{D^2,s}=\left|z_1-\bar{z}_2\right|^{-4(1+P_1^2)}\left|z_2-\bar{z}_2\right|^{2(P_1^2-P_2^2)}\\
&\times\int_0^\infty \frac{dP}{\pi}\mathcal{R}^s(P_1;P) \mathcal{R}^s(P_2;P) F\left(1+P_1^2,1+P_1^2,1+P_2^2,1+P_2^2;1+P^2\left|-\frac{(z_1-\bar{z}_1)(z_2-\bar{z}_2)}{|z_1-\bar{z}_2|^2}\right.\right),
\fe
where the relevant conformal block is again identified with a sphere 4-point holormophic Virasoro conformal block.

We can evaluate the Virasoro conformal blocks numerically as a truncated series in the elliptic nome using Zamolodchikov's recurrence relation, and evaluate (\ref{discopechann}) and (\ref{discbdrychann}) by numerically integrating the internal Liouville momentum. We have verified at random values of $z_1,z_2$ that the two channels agree numerically, with the relevant structure constants given as in (\ref{cppp}), (\ref{eq:FZZ1pt}), and (\ref{eq:bulkbdry}).

Next, we consider the correlation function of four boundary operators $\psi_{P_i}^{s,s}$ on the disc, subject to FZZT boundary condition.
Representing the disc as the upper half plane, we can put the four operators at 0, $x$, 1, $\infty$ on the real line, with $0<x<1$. The correlator can be represented in either the $s$-channel
\ie
&\left\langle \psi^{s,s}_{P_1}(0) \psi^{s,s}_{P_2}(x) \psi^{s,s}_{P_3}(1)\psi^{s,s}_{P_4}(\infty)\right\rangle_{D^2,s}=\\
&\int_0^\infty \frac{dP}{\pi}C^s(P_1,P_2,P)C^s(P_3,P_4,P) F\left(1+P_1^2,1+P_2^2,1+P_3^2,1+P_4^2;1+P^2\left|x\right.\right)
\fe
or the $t$-channel
\ie
&\left\langle \psi^{s,s}_{P_1}(0) \psi^{s,s}_{P_2}(x) \psi^{s,s}_{P_3}(1)\psi^{s,s}_{P_4}(\infty)\right\rangle_{D^2,s}=\\
&\int_0^\infty \frac{dP}{\pi}C^s(P_2,P_3,P)C^s(P_4,P_1,P) F\left(\left.1+P_2^2,1+P_3^2,1+P_4^2,1+P_1^2;1+P^2\right.|1-x\right).
\fe
Here for simplicity we have restricted to the case where the FZZT boundary conditions between each adjacent pairs of boundary operators are labeled by the same parameter $s$. 

We have checked numerically at random values of $x$ that the two-channels agree, using the boundary structure constant given in (\ref{eq:bdry3pt}).

\section{Fixing the normalization of boundary Liouville structure constants}
\label{sec:appnorm}

The normalization of the disc one-point function (\ref{eq:FZZ1pt}) is unambiguously fixed by the Cardy condition (\ref{cardyzz}), (\ref{cardyzzfzzt}). The crossing relation of Figure \ref{fig:disk2ptCA} and \ref{fig:disk2ptRR}, verified in Appendix \ref{sec:appcross}, further fixes the normalization of the disc bulk-boundary structure constant (\ref{eq:bulkbdry}). The crossing relation of the disc boundary four-point function, however, does not fix the overall normalization of the boundary structure constant (\ref{eq:bdry3pt}). In principle, the latter can be fixed by the crossing relations of correlation functions on the cylinder.

\begin{figure}[h!]
\centering
\subfloat[Bulk channel]{
{
\begin{tikzpicture}
\filldraw[color=black, fill=gray!40, thick] (0,0) circle (1);
\draw (-1,0) node[cross=4pt, thick] {};
\draw (-1,0) node[left] {$\psi_{P_1}^{s,s}$};
\draw (0,0) node[cross=4pt, thick] {};

\filldraw[color=black, fill=gray!40, thick] (3,0) circle (1);
\draw (4,0) node[cross=4pt, thick] {};
\draw (4,0) node[right] {$\psi_{P_2}^{s,s}$};
\draw (3,0) node[cross=4pt, thick] {};

\draw[snake it] (0,0) to (3,0);
\draw (1.5,0) node[above] {$V_{P}$};

\draw [thick] (5,0) -- (7,0);
\draw [thick] (7,0) -- (6.9,0.1);
\draw [thick] (7,0) -- (6.9,-0.1);
\draw (6,0) node[above] {$P_1\to i$};

\filldraw[color=black, fill=gray!40, thick] (9,0) circle (1);
\draw (9,0) node[cross=4pt, thick] {};

\filldraw[color=black, fill=gray!40, thick] (12,0) circle (1);
\draw (13,0) node[cross=4pt, thick] {};
\draw (13,0) node[right] {$\psi_{P_2}^{s,s}$};
\draw (12,0) node[cross=4pt, thick] {};

\draw[snake it] (9,0) to (12,0);
\draw (10.5,0) node[above] {$V_{P}$};

\end{tikzpicture}
\label{fig:ann2ptRR}
}}\\
\subfloat[Boundary channel]{
{
\begin{tikzpicture}
\filldraw[color=black, fill=gray!40, thick] (0,0) circle (1);
\draw (-1,0) node[cross=4pt, thick] {};
\draw (-1,0) node[left] {$\psi_{P_1}^{s,s}$};
\draw (0.707,0.707) node[cross=4pt, thick, rotate=45] {};
\draw (0.707,-0.707) node[cross=4pt, thick,rotate=45] {};

\draw[snake it] (0.707,0.707) to[bend left] (2.293,0.707);
\draw[snake it] (0.707,-0.707) to[bend right] (2.293,-0.707);
\draw (1.5,1) node[above] {$\psi_{P}^{s,s}$};
\draw (1.5,-1) node[below] {$\psi_{P'}^{s,s}$};

\filldraw[color=black, fill=gray!40, thick] (3,0) circle (1);
\draw (4,0) node[cross=4pt, thick] {};
\draw (4,0) node[right] {$\psi_{P_2}^{s,s}$};
\draw (2.293,0.707) node[cross=4pt, thick,rotate=45] {};
\draw (2.293,-0.707) node[cross=4pt, thick,rotate=45] {};

\draw [thick] (5,0) -- (7,0);
\draw [thick] (7,0) -- (6.9,0.1);
\draw [thick] (7,0) -- (6.9,-0.1);
\draw (6,0) node[above] {$P_1\to i$};

\filldraw[color=black, fill=gray!40, thick] (9,0) circle (1);
\draw (9.707,0.707) node[cross=4pt, thick, rotate=45] {};
\draw (9.707,-0.707) node[cross=4pt, thick,rotate=45] {};

\draw[snake it] (9.707,0.707) to[bend left] (11.293,0.707);
\draw[snake it] (9.707,-0.707) to[bend right] (11.293,-0.707);
\draw (10.5,1) node[above] {$\psi_{P}^{s,s}$};
\draw (10.5,-1) node[below] {$\psi_{P'}^{s,s}$};

\filldraw[color=black, fill=gray!40, thick] (12,0) circle (1);
\draw (13,0) node[cross=4pt, thick] {};
\draw (13,0) node[right] {$\psi_{P_2}^{s,s}$};
\draw (11.293,0.707) node[cross=4pt, thick,rotate=45] {};
\draw (11.293,-0.707) node[cross=4pt, thick,rotate=45] {};

\end{tikzpicture}
\label{fig:ann2ptCC}
}}
\caption{The two channels of the cylinder two-point function. In the limit $P_1\to i$, the cylinder two-point function becomes a cylinder 1-point function (up to a numerical constant).}
\label{fig:ann2pt}
\end{figure}
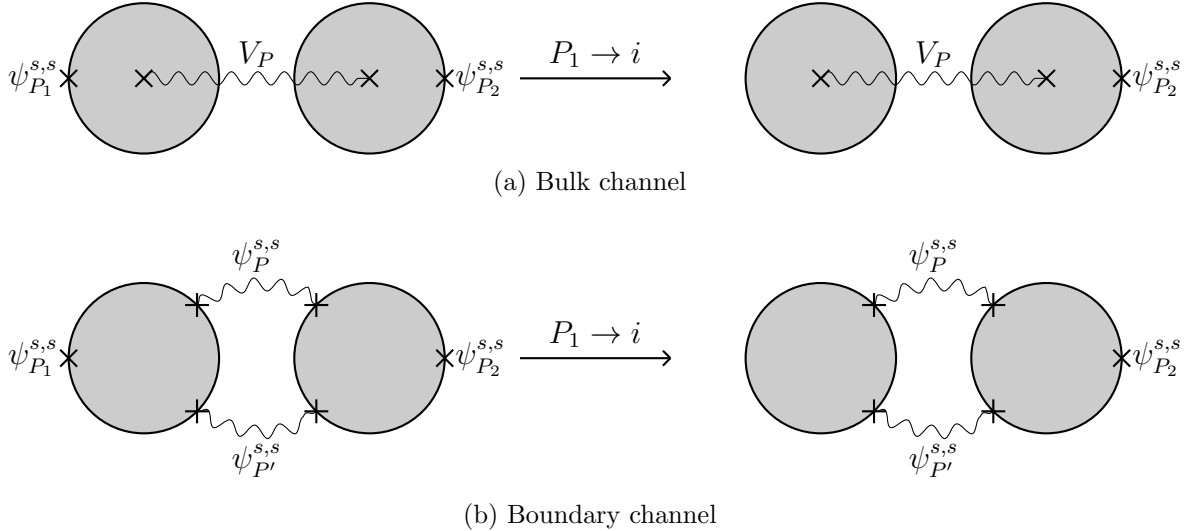

Let us consider the cylinder 2-point function, with operators $\psi_{P_1}^{s,s}$, $\psi_{P_2}^{s,s}$ inserted on the two boundaries respectively. The two different conformal block decompositions are represented by Figure \ref{fig:ann2pt}, which we refer to as the ``bulk channel" and the ``boundary channel", respectively. Our objective here is simply to assume the equivalence of the two channels, and use it to fix the normalization of the boundary 3-point function. This can be achieved by considering the analytic continuation $P_1\to i$, where the operator $\psi_{P_1}^{s,s}$ can effectively be ``erased", up to a normalization constant. In other words, the analytic continuation of $\psi_{P_1}^{s,s}$ to $P_1=i$ can be thought of as being proportional to the identity operator, even though the latter is not part of the boundary operator spectrum of the FZZT boundary condition.

In the bulk channel, 
upon analytic continuation to $P_1\to i$, the cylinder two-point function reduces to \cite{Hosomichi:2001xc}
\ie
\int_0^\infty &\frac{dP}{\pi} R^s(P;P_1)R^s(P;P_2) F^{\mathrm{cyl-2pt;bulk}}\left(\left.1+P_1^2,1+P_2^2;1+P^2\right|z\right)\\
&{\xrightarrow{P_1\to i}}2^\frac{3}{8}\pi^\frac{1}{4}\left(d^{s,s}(i)\right)^{-\frac{1}{2}}
\int_0^\infty \frac{dP}{\pi} \psi_s^{\mathrm{FZZT}}(P)R^s(P;P_2) F^{\mathrm{cyl-1pt;bulk}}\left(\left.1+P_2^2;1+P^2\right|z\right).
\fe
Here $F^{\mathrm{cyl-2pt;bulk}}$ and $F^{\mathrm{cyl-1pt;bulk}}$ are the bulk channel Virasoro conformal blocks for the cylinder 2-point and 1-point functions respectively. Their explicit expressions are not needed for our purpose. 

In the boundary channel, the same analytic continuation gives
\ie
\int_0^\infty & \frac{dP}{\pi} \frac{dP'}{\pi} C^{s,s,s}(P_1,P,P')C^{s,s,s}(P_2,P,P') F^{\mathrm{cyl-2pt;boundary}}\left(\left.1+P_1^2,1+P_2^2;1+P^2,1+P'^2\right|z\right)
\\
& {\xrightarrow{P_1\to i}}2^\frac{3}{8}\pi^\frac{5}{4} \left(d^{s,s}(i)\right)^{-\frac{1}{2}}
\int_0^\infty \frac{dP}{\pi} \frac{dP'}{\pi} \delta(P-P') C^{s,s,s}(P_2,P,P')F^{\mathrm{cyl-1pt;boundary}}\left(\left.1+P_2^2;1+P^2\right|z\right).
\fe
Demanding crossing equivalence of the two channels for both the cylinder 2-point function and 1-point function fixes the normalization of the boundary Liouville structure constants, as written in (\ref{eq:bulkbdry}) and (\ref{eq:bdry3pt}).


\section{Some details of numerical integration}
\label{sec:appnint}

In this appendix we describe some details on the numerical evaluation of the integrals in the Liouville momentum $P$ and modulus $x$ appearing in the worldsheet computation of long string amplitudes.

For the open $\to$ open + closed string amplitude, the moduli integral in $x$ must be regularized as in (\ref{eq:LLCreg}). Once the counter terms are taken into account, we can exchange the order of integration, perform the $x$-integral first and then integrate in Liouville momentum $P$. At each given value of $P$, we use Zamolodchikov's recursion relation to evaluate the conformal block as a function of $x$. The integration in $x$ in the vicinity $x=0$ is evaluated analytically by truncating the power series expansion of the integrand in $x$. Away from $x=0$, where the integrand is finite, we can evaluate the $x$-integral numerically for a set of sample $P$ values, interpolate in $P$ and then perform the $P$-integral numerically.

\begin{figure}[h!]
\centering
\includegraphics[width=0.7\textwidth]{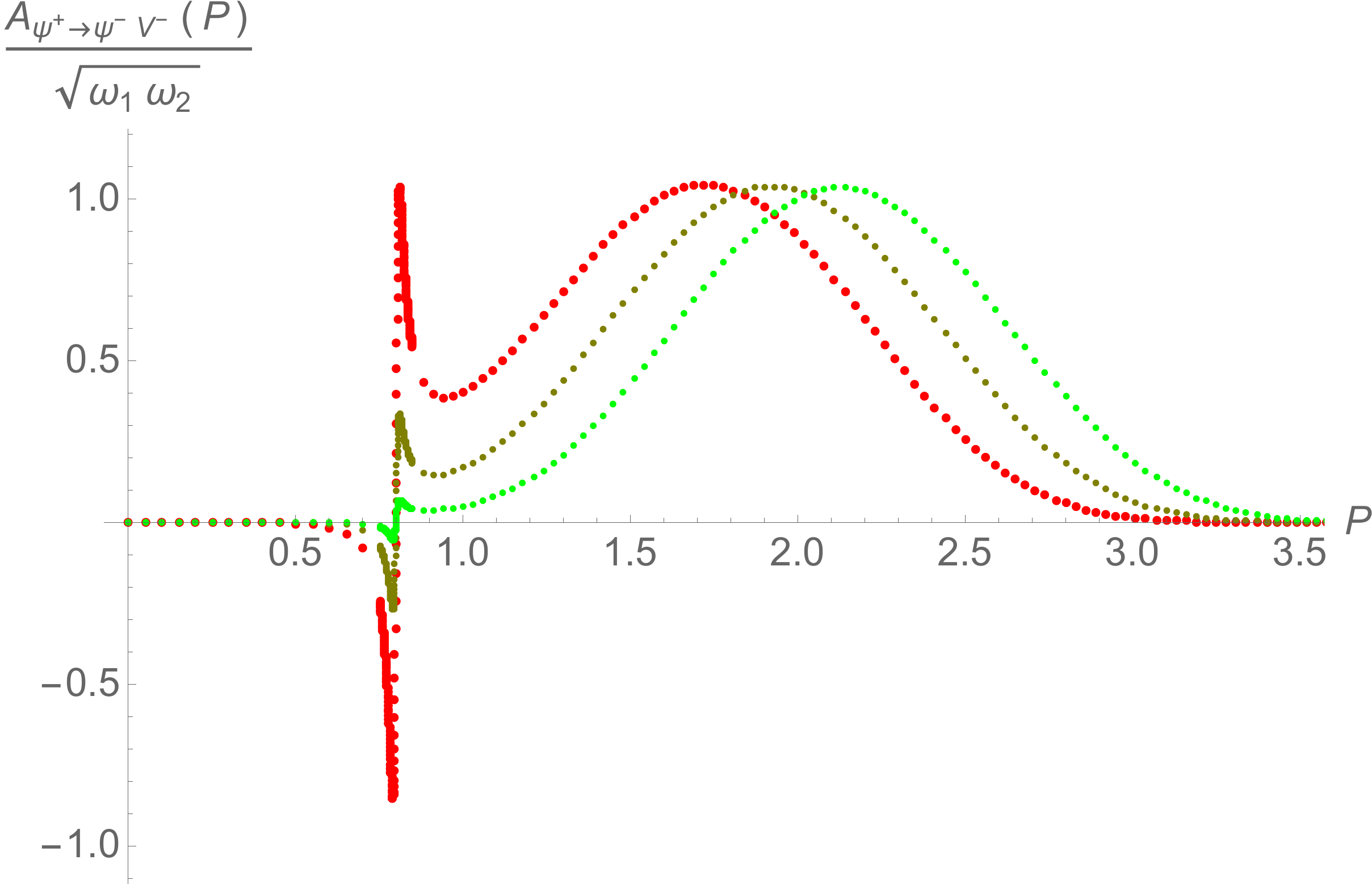}
\caption{A sample plot of the $P$-integrand in (\ref{eq:LLCreg}) for FZZT parameter $s=0.4,0.5,0.6$ (from red to green), after taking into account the normalization of the long string asymptotic state as in \ref{eq:lsamplitude}. The outgoing long string renormalized energy is taken to be $\epsilon_2=0.3$, and the closed string energy is $\omega_3=0.8+0.01 i$. We have performed the $x$-integral with the counter terms included for $P<\sqrt{\mathrm{Re}(\omega_3^2)}$. For sufficiently large $s$, the $P$-integral is dominated by the contribution near $P\sim \omega_1$, whereas the contribution from the $P$-integral up to $P\sim \sqrt{\mathrm{Re}(\omega_3^2)}$ (where the regulator is needed) becomes negligible.}
\label{fig:LLC_WS_regs}
\end{figure}


A typical plot of the $P$-integrand (after having performed the $x$-integral with regulator) is shown in Figure \ref{fig:LLC_WS_regs}. A small imaginary part of $\omega_3$ is included to regularize a potential divergence at $P=\sqrt{\mathrm{Re}(\omega_3^2)}$. A priori, due to the abrupt behavior of the integrand near $P=\sqrt{\mathrm{Re}(\omega_3^2)}$, finer sampling in $P$ is needed. However the contribution from this region is suppressed in the long string limit $s\to \infty$. In the end, it suffices to simply integrate over $P\gtrapprox \omega_3$ with real $\omega_3$ to obtain the long string amplitude.


The worldsheet computation of the long + long $\to$ long + long amplitude can be performed similarly. The diagrams come in two types: ones like Figure \ref{fig:LLLL}a that requires regulator up to $P<\sqrt{\mathrm{Re}(\omega_1+\omega_2)^2}\sim 2s$, and ones like Figure  \ref{fig:LLLL}b that requires regulator up to $\sqrt{\mathrm{Re}(\omega_1-\omega_3)^2}$. The former requires evaluating the conformal blocks to high precision and the result is a delicate cancelation in the long string limit. In fact, as explained in section \ref{sec:WSLLLL}, such diagrams are exponentially suppressed at large $s$ and do not contribute to the long string amplitude. The surviving diagrams, of the type Figure  \ref{fig:LLLL}b, can be treated again by restricting the lower range of the $P$-integral to avoid the need for regulators, in the long string limit.


\section{Resonance computation of open $\to$ open + closed string amplitude}
\label{sec:appres}

In this section we give some details of the derivation of the resonance amplitude (\ref{eq:LLCres}),
defined by analytically continuing the open $\to$ open + closed string amplitude up to imaginary closed string energy $\omega_3=i$, at fixed outgoing open string energy $\omega_2$. More precisely, we will set $\omega_3=i(1-\epsilon)$ and take the limit $\epsilon\to0^+$. In this limit, the bulk-boundary 2-point goes to zero, but this is compensated by a divergence in the $P$-integral as $P\to 0$ coming from both the bulk-boundary 2-point and the boundary 3-point functions. In this sense, we can say that the resonance amplitude localizes to $P=0$.



Let us first examine the bulk-boundary structure constant ${\cal R}^s(P_3; P)$, given by (\ref{eq:bulkbdry}) with the appropriate relabeling of Liouville momenta, under the analytic continuation of the bulk vertex operator to $P_3 = {\omega_3\over 2} = i{1-\epsilon\over 2}$. We may equivalently take the $t$-contour in (\ref{eq:bulkbdry}) to be $C_\delta=\mathbb{R}+i\delta$ with ${\epsilon\over 2}<\delta<{1\over 2}$, and keep track of poles of the integrand crossing the contour.  
One can verify that the integral over $C_\delta$ as in the second line of (\ref{eq:bulkbdry}) remains finite in the $\epsilon\to 0$ limit at any $P$. The prefactor which contains $(S(P_3))^{-1/2}$ vanishes like $\epsilon$ in this limit.

However, in such an analytic continuation, a pole of the integrand of (\ref{eq:bulkbdry}) has crossed the contour $C_\delta$ to the location $t=\frac{1}{2}(\omega_3+P)+\frac{i}{2}$. Its residue contribution is
\ie{}
& 2^{\frac{5}{8}}\pi^\frac{3}{4} \left(S\left(\frac{\omega_3}{2}\right)\right)^{-1/2}\left(d^{s,s}(P)\right)^{-1/2}\frac{\Gamma_1(1-iP)^3\Gamma_1(1-i(\omega_3+P))\Gamma_1(1+i(\omega_3-P))}{\Gamma_1(2)\Gamma(1+iP)\Gamma_1(-2iP)\Gamma_1(2+i\omega_3)\Gamma_1(-i\omega_3)}\\
&~~~ \times  e^{-2\pi s(1+i(\omega_3+P))} \frac{S_1(1+i(\omega_3+P))}{S_1(1-iP)S_1(2+i\omega_3)}.
\label{eq:bulkbdryres}
\fe
For generic $P$, the factor in the second line is finite and thus the residue contribution to ${\cal R}^s(P_3; P)$ vanishes as $\epsilon\to0$. However, near $P=0$, the residue contribution takes the form
\ie
-2^\frac{5}{8}\pi^\frac{3}{4} \left(-d^{s,s}(P)\right)^{-1/2} \epsilon \frac{\Gamma_1(\epsilon-iP)}{\Gamma_1(-2iP)}S_1(\epsilon+i P)\to \frac{2^\frac{5}{8}}{\pi^\frac{1}{4}}i \frac{\epsilon P}{P^2+\epsilon^2}\left(-d^{s,s}(P)\right)^{-1/2}.
\label{eq:bdrybulkres}
\fe


Next consider the boundary structure constant $C^{s,s,s}(P,\omega_1,\omega_2)$, given by (\ref{eq:bdry3pt}) with the appropriate relabeling of Liouville momenta, analytically continued to $\omega_1=\omega_2+i(1-\epsilon)$. One can check that the contour integral along the real line as in (\ref{eq:bdry3pt}) gives a finite result for any $P$. However, in the analytic continuation, a pole has crossed the contour, giving the residue contribution
\ie\label{cpppresid}
& 2^\frac{3}{8}\pi^\frac{5}{4}\frac{\left(d^{s,s}(P)\right)^\frac{1}{2}}{\left(d^{s,s}(\omega_1)\right)^\frac{1}{2}\left(d^{s,s}(\omega_2)\right)^\frac{1}{2}}\frac{S_1(1+iP)S_1(1+i(P-2s))}{S_1(1+i\omega_2)S_1(1+i(\omega_2-2s))}\\
&\times \frac{\Gamma_1(1-i(\omega_1+\omega_2+P))\Gamma_1(1+i(\omega_2+P-\omega_1))\Gamma_1(1+i(\omega_2-P-\omega_1))\Gamma_1(1+i(P-\omega_2-\omega_1))}{\Gamma_1(2)\Gamma_1(2iP)\Gamma_1(-2i\omega_2)\Gamma_1(-2i\omega_1)}\\
&\times  2\pi i\mathrm{Res}_{t=i-\omega_1}\prod_{k=1}^4\frac{S_1(U_k+i t)}{S_1(V_k+it )},
\fe
where $U_k$ and $V_k$ are defined as below (\ref{eq:bdry3pt}), again with the appropriate relabeling of Liouville momenta.
(\ref{cpppresid}) is finite away from $P=0$. Together with the vanishing ${\cal R}^s(P_3; P)$ in the $\epsilon\to 0$ limit, we learn that the resonance amplitude does not receive contribution from the $P$-integral away from $P=0$.

In the limit of small $P$ and $\epsilon$, (\ref{cpppresid}) reduces to
\ie
2^{-\frac{5}{8}}\pi^\frac{1}{4}&\frac{P}{\epsilon^2+P^2}\frac{\sinh(2\pi\omega_2)}{\sinh(\pi\omega_2)}\left[\frac{(2i\omega_2)(-1+2i\omega_2)^2(-2+2i\omega_2)}{\sinh(2\pi s+\pi\omega_2)\sinh(\pi\omega_2-2\pi s)}\right]^\frac{1}{2}.
\label{eq:bdry3ptres}
\fe
Combining (\ref{eq:bdry3ptres}) and (\ref{eq:bulkbdryres}), we can perform the $P$-integral near $P=0$ and conclude that in the $\epsilon\to 0$ limit,
\ie\label{pintlocal}
&\int_0^\infty \frac{dP}{\pi}R^s\left(\frac{\omega_3}{2};P\right)C^{s,s,s}(P,\omega_1,\omega_2)i^{-\omega_1^2+\omega_2^2-1-P^2}F\left(1+\omega_1^2,1+\omega_2^2,1+\frac{\omega_3^2}{4},1+\frac{\omega_3^2}{4};1+P^2|\eta\right)\\
&\to -\frac{i^{-\omega_1^2+\omega_2^2}}{4}\frac{\sinh(2\pi\omega_2)}{\sinh(\pi\omega_2)}\left[\frac{(2i\omega_2)(-1+2i\omega_2)^2(-2+2i\omega_2)}{\sinh(2\pi s+\pi\omega_2)\sinh(\pi\omega_2-2\pi s)}\right]^\frac{1}{2}F\left(1+\omega_1^2,1+\omega_2^2,1+\frac{\omega_3^2}{4},1+\frac{\omega_3^2}{4};1|\eta\right).
\fe
That is, the boundary Liouville correlator of interest is proportional to a single Virasoro conformal block, which also coincides with a 4-point function in the linear dilaton CFT,
%
\ie
F\left(\left.1+(\omega_2+i)^2,1+\omega_2^2,\frac{3}{4},\frac{3}{4};1\right|\eta\right)=\eta^{-2i\omega_2-2\omega_2^2}(1-\eta)^{-i\omega_2}.
\fe
Including the free boson correlator in (\ref{eq:LLCfull}), and the $x$-dependent prefactors as in (\ref{eq:LLCLiouv}), we end up with the moduli integral
\ie
\int_0^\infty dx\left|x-\frac{i}{2}\right|^{-2}=\pi.
\fe
Putting all this together, with the prefactors in (\ref{pintlocal}), we arrive at (\ref{eq:LLCres}).

\section{An alternative parameterization of collective excitations of the fermi surface}
\label{pparasec}

In this appendix we describe an alternative parametrization of the collective excitations of the fermi sea in the singlet sector of the $c=1$ matrix quantum mechanics. That is, we view the collective fields as functions of the momentum $p$ rather than position $\lambda$ of the single-fermion/eigenvalue phase space. While not strictly needed in our calculations of scattering amplitudes so far, the $p$-parameterization has the advantage that it avoids dealing with a collective field Hamiltonian that is singular at the ``tip" $\tau=0$ of the fermi surface, thereby eliminating the need for the regularization described in section \ref{singletsection}.


To derive the collective field Hamiltonian in the $p$-parameterization, we need to introduce an IR regulator at large $\lambda$ to make the fermi surface compact. We will do so by adding a quartic term to the single fermion Hamiltonian, now written as
\ie
H_i = \frac{p^2}{2} - \frac{\lambda_i^2}{2} + \alpha \lambda_i^4,
\fe
where $\alpha$ is a positive parameter that will be taken to zero in the end. 

The total Hamiltonian of the system can be written as
\ie
\label{eq:Hcollp}
H&=\int \frac{d\lambda dp}{2\pi}\epsilon\theta(\epsilon_F-\epsilon)+\mu N\\
&=\int \frac{dp}{2\pi}\left[ \alpha\left(\frac{\lambda_+(p)^5}{5}-\frac{\lambda_-(p)^5}{5}\right)-\left(\frac{\lambda_+(p)^3}{6}-\frac{\lambda_-(p)^3}{6}\right)+\left(\mu+\frac{p^2}{2}\right)\left(\lambda_+(p)-\lambda_-(p)\right)\right],
\fe
where $\lambda_\pm(p)$ are the maximal/minimal values of the eigenvalue $\lambda$ of given $p$ in the fermi sea, related to the momentum density $\Pi(p)$ and eigenvalue density $\phi(p)$ by
\ie
\Pi(p )&\equiv\sum_{i=1}^N\delta(p-p_i)=\frac{1}{2\pi}\int_{\lambda_-(p )}^{\lambda_+(p)}d\lambda=\frac{\lambda_+(p )-\lambda_-(p )}{2\pi},\\
\phi(p)&\equiv\sum_{i=1}^N\lambda_i \delta(p-p_i)=\frac{1}{2\pi}\int_{\lambda_-(p )}^{\lambda_+(p)}\lambda d\lambda=\frac{\lambda_+(p )^2-\lambda_-(p )^2}{4\pi}.
\fe
$\Pi(p)$ and $\phi(p)$ obey the Poisson bracket $\{\Pi(p),\phi(p')\}_{\rm P}=\partial_p(p-p')\Pi(p)$.

The ground state of the fermi sea corresponds to the profile 
\ie
\lambda^0_{\pm}(p)=\frac{1}{2} \sqrt{\frac{1}{\alpha}\pm\frac{\sqrt{1-8 p^2\alpha-16\alpha\mu}}{\alpha}},
\fe
whereas fluctuations of the fermi surface can be parameterized as
\ie
\lambda_\pm(p)=\lambda_\pm^0(p)+\sqrt{\pi}\left(\Pi_\eta\pm\partial_p\eta\right).
\fe
Here $\eta(p)$ is the collective field in $p$-parameterization, and $\Pi_\eta$ its conjugate canonical momentum density.
One can substitute this into (\ref{eq:Hcollp}), pass to the $\tau$ coordinate defined by $p=\sqrt{2\mu}\sinh\tau$ (which now ranges over the entire real line), and derive the action $S[\eta]$. It is then straightforward to take $\alpha\to 0$ at the level of the action, resulting in
\ie
S[\eta] =\int_{-\infty}^{\infty} d\tau \left(-\dot\eta\partial_\tau\eta-\left(\partial_\tau\eta\right)^2+\frac{\sqrt{\pi}}{3\mu\cosh^2\tau}\left(\partial_\tau\eta\right)^3\right)
\label{eq:CollFT_lag}
\fe
describing a relativistic right-moving massless field with cubic interaction. Notice that the interaction term is perfectly regular at $\tau= 0$, and we can, for instance, recover from it the same tree level $1\to 2$ amplitude of collective-field/closed-strings ${\cal A}_{1\to 2} = {1\over \mu} i\omega \omega_1\omega_2$.

\bibliographystyle{JHEP}
\bibliography{longstringdraft}

\end{document}